\documentclass[prb,twocolumn,showpacs]{revtex4}
\pdfoutput=1
\usepackage{amssymb,latexsym}
\usepackage{amsmath,amsbsy,bbm}
\usepackage{ifpdf}
\usepackage{epsfig,bm}
\usepackage{graphicx,comment}
\usepackage{color}
\usepackage{soul}
\usepackage{mathtools}
\usepackage[normalem]{ulem}
\begin{document}

\title{Machine learning phases and criticalities without using real data for training}
\author{D.-R. Tan}
\affiliation{Department of Physics, National Taiwan Normal University,
88, Sec.4, Ting-Chou Rd., Taipei 116, Taiwan}
\author{F.-J. Jiang}
\email[]{fjjiang@ntnu.edu.tw}
\affiliation{Department of Physics, National Taiwan Normal University,
88, Sec.4, Ting-Chou Rd., Taipei 116, Taiwan}

\begin{abstract}
  We study the phase transitions of three-dimensional (3D) classical $O(3)$ model and
  two-dimensional (2D) classical XY model, as well as both the quantum phase
  transitions of 2D and 3D dimerized spin-1/2 antiferromagnets, using
  the technique of supervised neural network (NN). Moreover, unlike the conventional
  approaches commonly used in the literature, the training sets employed
  in our investigation are neither the theoretical nor the real configurations of
  the considered systems. Remarkably, with such an unconventional set up of the training
  stage in conjunction with some semi-experimental finite-size scaling formulas, the associated critical
  points determined by the NN method agree well with the established results in
  the literature. The outcomes obtained here imply that
  certain unconventional training strategies, like the one used in this study, are not only
  cost-effective in computation, but are also applicable for a wild range of physical systems.

\end{abstract}


\maketitle

\section{Introduction}

The applications of artificial intelligence (AI) methods and techniques
to the studies of many-body systems have recently inspired the communities of physics, applied physics and
physical chemistry.
Moreover, many important and exciting achievements have been obtained using the AI approach in
the last a few years \cite{Rup12,Sny12,Mon13,Pil13,Mer14,Sch14,Li15,Baldi:2014pta,Mnih:2015jgp,Lan16,Lee16,Searcy:2015apa,Castro:2015rrx,Baldi:2016fzo,Baldi:2016fql,Tor16,Wan16,Att16,Oht16,Hoyle:2015yha,Car16,Tro16,Wu17,Bro16,Chn16,Barnard:2016qma,Tan16,Tubiana:2016zpw,Nie16,Mott:2017xdb,Liu16,Tam17,Xu16,Wan17,Zevin:2016qwy,Liu17,Liu17.1,Che17,Nag17,Kol17,Den17,Pon17,Kasieczka:2017nvn,Zha17,Zha17.1,Hu17,Li18,Chn18,Lu18,George:2016hay,Bea18,Pang:2016vdc,Sha18,Art18,But18,Bar18,Zha18,Gra18,Butter:2017cot,Gao18,Zha19,Gre19,Ren:2017ymm,Don19,Cavaglia:2018xjq,Dav19,Yoo19,Conangla:2018nnn,Fluri:2019qtp,Can19,Li19,Cha19,Lia19,Zhu19,Meh19,Sch19,Car19,Oht20,Has20,Jin20,Larkoski:2017jix,Han:2019wue,Tan20.1}. Among these achievements, first principles calculations of
properties
of materials and analyzing the signals from colliders in high energy
physics are two such notable examples. Yet another significant
accomplishment is the success of investigating critical phenomena
using both the supervised and unsupervised neural networks (NN).

By employing the dedicated convolutional neural network techniques (CNN)
which can capture certain characteristics of the studied models,
it has been demonstrated that the phase transitions associated with many classical and quantum systems, including the Ising model, the XY models,
as well as the Hubbard model have been studied with various extent of satisfaction.
Because of these numerous successful examples mentioned, it is optimistically believed that with the ideas of AI one may be able to 
uncover features of certain systems that cannot be obtained by
the conventional methods. Even those days, seeking devoted AI techniques to
surpass the success that the traditional approaches
can reach is still vigorous.  

The standard procedure, i.e., the most considered scheme, of
investigating the phase
transitions of physical models by supervised NN
consists of three steps \cite{Car16,Meh19,Car19},
namely the training, the validation,
and the testing stages. Among these three stages, the training
is the most flexible one and various strategies have
be used for this step \cite{Car16,Nie16,Li18,Tan20.1}.
Typically real configurations of the studied systems obtained
from certain numerical methods are employed as the training sets.
In addition, the training has been applied to various chosen
temperatures $T$ (relevant parameter) across the transition temperature $T_c$
(critical point). This indicates that in principle $T_c$ (or the critical
point) should be known in advance before one can employ the NN techniques.
Such a training approach has led to success in studying the critical
phenomena associated with several many-body systems such
as the Ising and the Hubbard models. Other schemes for which
the locations of the critical points are not required
are introduced as well \cite{Li18,Rod19,Tan20.1,Sin20}.
For instance, the method of using the
theoretical ground state configurations in the ordered phase
as the training sets are demonstrated to be valid for ferromagnetic and
antiferromagnetic Potts models. 
For the readers who are interested in the details of these training processes, 
see Refs.~\cite{Car16,Li18,Meh19,Car19,Tan20.1}.

The strategy of considering the theoretical ground states in the ordered phase
as the training sets requires only one training and the
knowledge of the associated critical point(s) is not needed. 
This approach has been applied
to both the ferromagnetic and the antiferromagnetic Potts models, and
the obtained outcomes show that the idea is effective \cite{Li18,Tan20.1}.
In particular, numerical evidence strongly suggests that with this method,
the computational demanding for the training stage is tremendously reduced,
and its applicability is broad. 

Despite the NN results of estimating the critical points associated with the Potts models,
using the
method of considering the ground state configurations
in the ordered phase as the training sets, is impressive, an interesting question arises.
Specifically, is this approach applicable for studying the zero temperature phase transitions
of quantum spin systems, as well as the phase transitions of models with continuous variables
such as the classical $O(3)$ model?

To answer the crucial question outlined in the previous paragraph, here we study the phase transitions
of three-dimensional (3D) classical $O(3)$ (Heisenberg) model and two-dimensional (2D) classical XY model,
as well as the quantum phase transitions
of 2D and 3D dimerized spin-1/2 antiferromagnets, using the simplest deep learning neural network,
namely the multilayer perceptron (MLP). In particular, unlike the conventional or the unconventional
training procedures introduced previously, in this investigation, following the idea of using
the theoretical ground states in the ordered phase as the training sets, we have adopted an alternative
strategy for the training. In particular, the training sets employed here belong to neither
the theoretical nor the real configurations of the considered systems. 

The motivation for using the simplest deep learning NN in the study is that 
whether a NN idea is valid or not should not depend on the detailed infrastructure of the built NN.
Hence a MLP made up of only three layers are employed here.  
One can definitely considered a more complicated (and dedicated as well) NN
such as CNN for the associated investigations.
This will be left for future work.

Remarkably, even using the extraordinary training sets mentioned above and some semi-experimental finite-size scaling
(which will be introduced later), the constructed MLP can effectively detect
the critical points of all the studied classical and quantum physical systems. The intriguing
outcomes obtained here strongly suggest that the approach of investigating the targeted
physical systems before employing any objects for the training, such as those done here and in \cite{Li18,Tan20.1},
is not only cost-effective in computation, but also leads to accurate determination of the associated critical points.
Finally, it is amazing that the simple procedure described here is not only valid for studying the
phase transitions associated with spontaneous symmetry breaking (SSB), but also works for those related to topology.

This paper is organized as follows. After the introduction,
the studied microscopic models and the employed NN are 
described. In particular, the NN training sets and labels are introduced 
thoroughly. Following this the resulting numerical results determined by
applying
the NN techniques are presented. 
Finally, a section concludes our investigation.

\begin{figure}
\begin{center}
\hbox{
\includegraphics[width=0.22\textwidth]{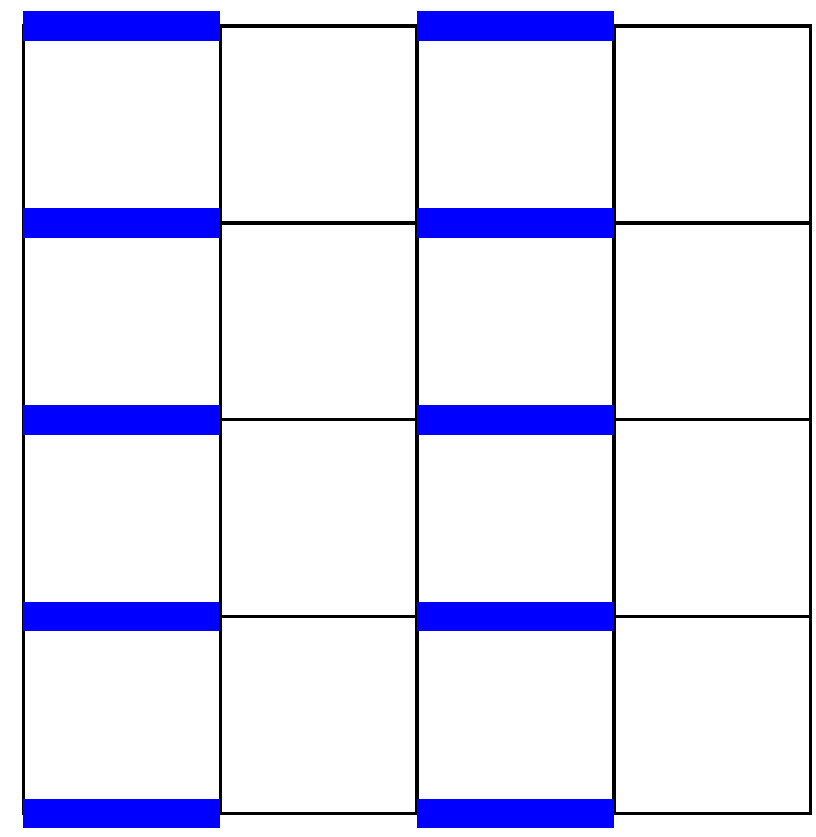}~~~~
\includegraphics[width=0.24\textwidth]{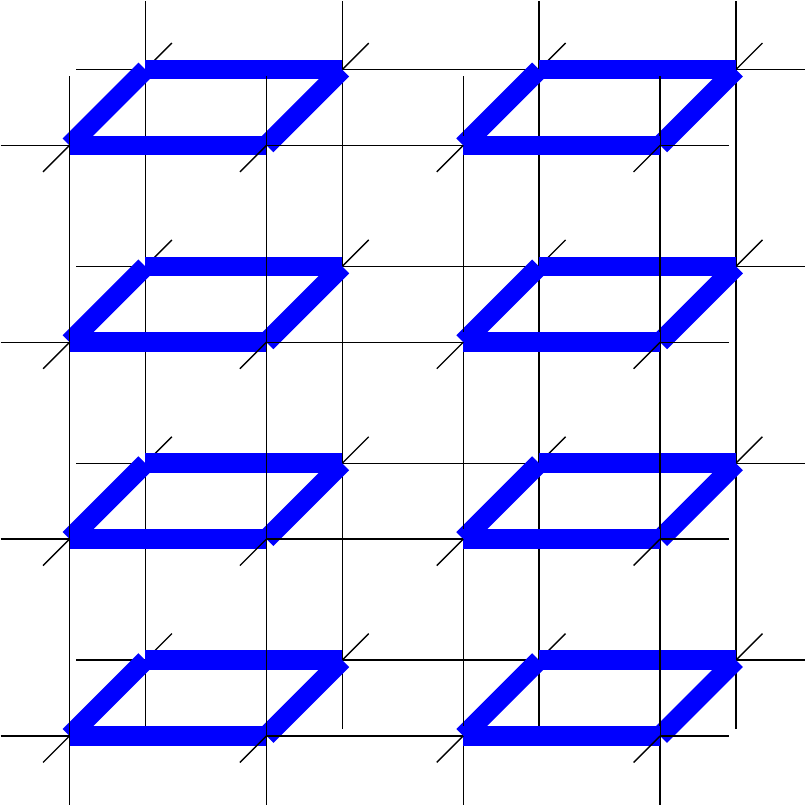}
}
\end{center}\vskip-0.7cm
\caption{The studied dimerized quantum antiferromagnetic Heisenberg models:
  2D ladder (left) and 3D plaquette (right) models. The bold and thin bonds
  shown in both sub-figures represent $J'$ and $J$ couplings, respectively.}
\label{model_quantum}
\end{figure}

\begin{figure*}
  \begin{center}
      \includegraphics[width=0.9\textwidth]{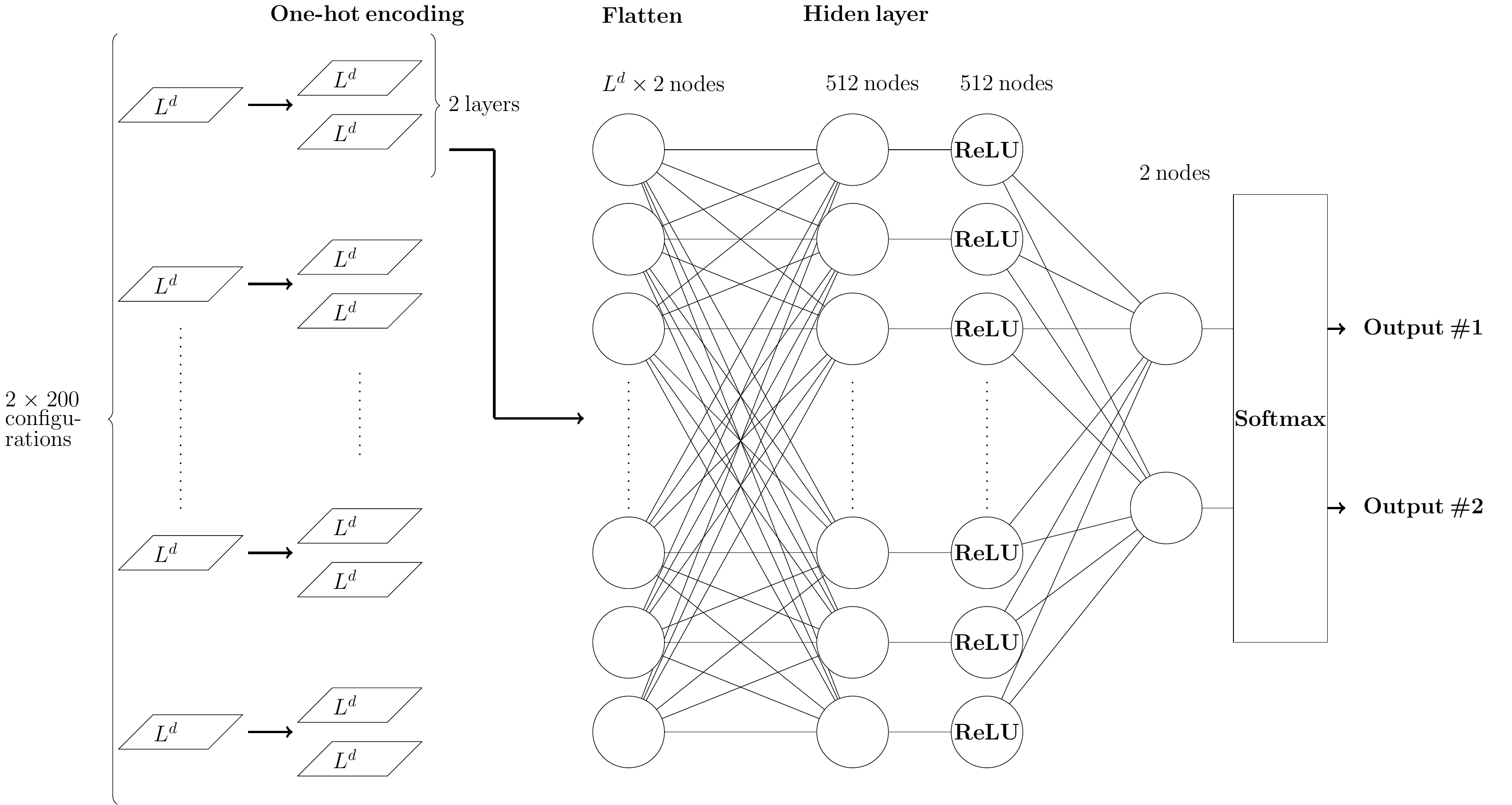}
\end{center}\vskip-0.7cm
\caption{The NN (MLP), which consists of one input layer, one hidden layer,
  and one output layer, used here and in Ref.~\cite{Tan20.1}.
  In the figure $d$ is the dimensionality of the considered system. In addition,
  the objects in the input layer are
  made up of 200 copies of only two configurations for all the studied models.
  Finally, there are 512 (or 1024) nodes in the hidden layer and each of these
  nodes is independently connected to every object in the input layer. Before
  each training object is connected to the nodes in the hidden layer,
  the steps of one-hot encoding and flatten are applied. The activation
  functions (ReLU and softmax) and where they are employed are demonstrated
  explicitly. For all the considered systems, the output layers consist of two elements.}  
\label{MLP}
\end{figure*}

\section{The microscopic models and observables}

\subsection{The 3D classical $O(3)$ (Heisenberg) and 2D classical XY model}

The Hamiltonian $H_{O(3)}$ of the 3D classical $O(3)$ (Heisenberg) model on a cubical lattice
considered in our study is
given by 
\begin{equation}
\beta H_{O(3)} = -\beta \sum_{\left< ij\right>} \vec{s}_i\cdot\vec{s}_j,
\label{eqn1}
\end{equation}
where $\beta$ is the inverse temperature and $\left< ij \right>$ stands for 
the nearest neighboring sites $i$ and $j$. In addition, in Eq.~(\ref{eqn1})
$\vec{s}_i$ is a unit vector belonging to a 3D sphere $S^3$ and is located at
site $i$. 

Starting with an extremely low temperature, as $T$ rises,
the classical $O(3)$ system will undergo a phase
transition from an ordered phase, where majority of the unit vectors point toward the same direction, to a
disordered phase for which these mentioned vectors are oriented randomly. Relevant observables used here
to signal out the phenomenon of this phase transition are the first and the second Binder ratios ($Q_1$ and $Q_2$) defined by
\begin{eqnarray}
  Q_1 &=& \langle |m| \rangle^2/ \langle m^2\rangle, \\
  Q_2 &=& \langle m^2\rangle^2 / \langle m^4 \rangle,
  \end{eqnarray}
where $m = \frac{1}{L^3}\sum_i \vec{s}_i$ and $L$ is the linear box size of the system \cite{Bin81}. 

The Hamiltonian of the 2D classical XY model on the square lattice has the same expression as $H_{O(3)}$, except
that the corresponding unit vector $\vec{s}_i$ at site $i$ belongs to a (2D) circle instead of a 3D sphere. 

\subsection{The 2D and 3D dimerized quantum antiferromagnetic Heisenberg models}

The 2D and 3D dimerized quantum antiferromagnetic Heisenberg model share a similar
form of Hamiltonian given as

\begin{eqnarray}
H = \sum_{\langle i, j \rangle}J_{ij} \vec{S}_i \cdot \vec{S_j},
\end{eqnarray}
where again $\langle i, j\rangle$ stands for the nearest neighboring sites $i$ and $j$,
$J_{ij} > 0$ is the associated antiferromagnetic coupling (bond) connecting $i$ and
$j$, and $\vec{S}_i$ is the spin-1/2
operator located at $i$. The cartoon representation, in particular the spatial arrangement of
the antiferromagnetic couplings, of the studied models are shown in fig.~\ref{model_quantum}
(In this study, these quantum spin models will be called 2D ladder and 3D plaquette models
if no confusion arises).
From the figure one sees that, as the ratios $J'/J$ (of both models) being tuned,
quantum phase transitions from ordered to disordered states will
take place in these models when $g\coloneqq J'/J$
exceed certain values $g_c$. Relevant observables
considered in our investigation for
studying the quantum phase transitions are again the first and the second Binder ratios
described above. For the studied spin-1/2 systems, $Q_1$ and $Q_2$ have the following definitions
\begin{eqnarray}
  Q_1 &=& \langle |M_s| \rangle^2/ \langle M_s^2\rangle, \\
  Q_2 &=& \langle M_s^2\rangle^2 / \langle M_s^4 \rangle,\\
  M_s &=& \frac{1}{L^d} \sum_{i} (-1)^{i_1 + i_2} S_i^z, 
  \end{eqnarray}
 here $d$ is the dimensionality of the studied models.

 These mentioned $g_c$ of the quantum spin systems, as well as the $T_c$ of the 3D classical $O(3)$ and 2D classical XY models introduced previously, have been calculated with
 high accuracy in the literature \cite{Hol92,Cam02,Has05,San10,Tan18}.

\section{The constructed supervised Neural Networks}

In this section, we will review the supervised NN, namely the multilayer perceptron (MLP) used in our study.
The employed training sets and the associated labels for the studied models will 
be described as well.

\subsection{The built multilayer perceptron (MLP)}

The MLP used in our investigation is already detailed in Ref.~\cite{Tan20.1}. 
Specifically, using the NN library keras \cite{kera}, we construct a supervised NN
which consists of only one input layer, one hidden layer of 512 (or 1024) 
independent nodes, and one output layer. In addition, The algorithm, optimizer, and
loss function considered in our calculations are the minibatch, the adam, and the
categorical cross entropy, respectively.
To avoid overfitting, we also apply $L_2$ regularization at various stages.
The activation functions employed here are ReLU and softmax.
The details of the constructed MLP, including the steps of
one-hot encoding and flatten (and how these two processes work)
are shown in fig.~\ref{MLP} and are available in Ref.~\cite{Tan20.1}.

Finally, for the three studied models, results calculated using 10 sets of random seeds
are all taken into account when presenting the final outcomes. We would like
to point out that in the testing stage, each of these 10 calculations uses
the same set of configurations produced from the Monte Carlo simulations.
Later we will come back to this and make a comment about it.

\subsection{Training set and output labels for the 3D classical $O(3)$ and 2D classical XY models}

Regarding the training set employed in the calculations, instead of using
real configurations obtained from simulations or
the theoretical ground states in the ordered phase of
the considered system,
here we use a slightly different alternative. Specifically, to train
the NN on a $L$ by $L$ by $L$ cubical lattice for the 3D classical $O(3)$ model,
the training set consists of only two configurations.
In addition, 0 is assigned to 
every site of one configuration and the other configuration is made up
by giving each of its sites the value of 1.
As a result, the output labels are
the vectors of $(1,0)$ and $(0,1)$. 
The same configurations and labels are employed for the 2D classical XY model as well.
The motivation for considering such training sets will be explained in
next subsection.

\subsection{The expected output vectors for the 3D classical $O(3)$ and 2D classical XY models at various $T$}

It should be pointed out that an $O(3)$
configuration is specified completely by the associated two parameters, namely $\theta$ and $\psi$
at each site of the underlying cubical lattice. At an extremely low temperature $T$,
all the unit vectors of a $O(3)$ configuration point toward a particular direction.
Under such a circumstance, $\psi$ mod $\pi$ is either 0 or 1 for every unit vector (of an $O(3)$ configuration). The employed training
set described in the previous subsection is motivated by this observation.
As a result, the magnitude $R$ of the output vector
for a ground state $O(3)$ configuration is 1. When the temperature rises,
one expects that $R$ diminishes with $T$ and for $T \ge T_c$, $R$ takes its
possible minimum value $1/\sqrt{2}$. Consequently, studying the magnitude of the
NN output vectors as a function of $T$ can reveal certain relevant information of $T_c$.

Here we would like to emphasize the fact that $\psi$ rather than $\theta$ is considered
in our investigation.
This is because for any two given fixed values of $\psi$, their related arc length
on the 3D unit sphere are the same. For $\theta$, this is not the case.
Therefore, with $\psi$ one should arrive at more accurate outcomes. 

The same scenario described above for the 3D classical $O(3)$ model applies to the 2D classical
XY model as well. 

\subsection{Training set and output labels for the 2D and 3D quantum spin models}

For the 2D and 3D dimerized quantum antiferromagnetic Heisenberg models investigated here,
their associated classical ground state configurations (in the order phase) are adopted
as the training set. Specifically, the training set for each of these two models consists
of two configurations. Moreover, the spin value of every lattice site is either
1 or -1 and they are arranged alternatively. In other words, for a site which has
a spin value 1 (-1), the spin values for all of its  nearest neighbor sites are
-1 (1). With such set up of the training sets, the employed output vectors
should be $(1,0)$ and $(0,1)$ naturally.

We would like to emphasize the fact that the training sets considered for
the studied 2D and 3D quantum spin models are not even among any of the
possible ground state configurations of these two systems.

\subsection{The expected output vectors for the 2D and 3D dimerized quantum spin models at various $g$}

Due to quantum fluctuations, it is not possible to assign any definite spin configurations
for these investigated quantum spin models when $g = 1$ and $g > g_c$. Therefore,
how the corresponding output vectors behave with respect to the dimerized strength $g$
will be treated classically here. Consequently, $R$ should be 1 and $1/\sqrt{2}$
for $g = 1$ and $g \ge g_c$, respectively. As we will demonstrate shortly,
the $R$ (magnitude) of the outputs associated with the NN studies of these quantum spin systems
follow these rules (i.e., the values of $R$ are 1 and $1/\sqrt{2}$
for $g = 1$ and $g > g_c$, respectively) in a satisfactory manner, hence lead to fairly good estimations
of the critical points.

\section{The numerical results}

The configurations associated with the considered systems, namely the 3D (classical) $O(3)$, the 2D classical XY, the 3D plaquette,
as well as the 2D ladder models are generated by the Wolff and
the stochastic series expansion (SSE) algorithms \cite{Wol89,San99,San10}.
In addition, for each of the studied model, the corresponding configurations are
recorded once in (at least) every two thousand Monte Carlo sweeps after the thermalization,
and at least one thousand configurations are produced. These spin compositions are then
used for the calculations of NN. A semi-experimental finite-size scaling, which
is adopted to estimate the critical points, will be
introduced as well in this section.

\subsection{Results of 3D classical $O(3)$ model}
\begin{figure}
\begin{center}
\includegraphics[width=0.5\textwidth]{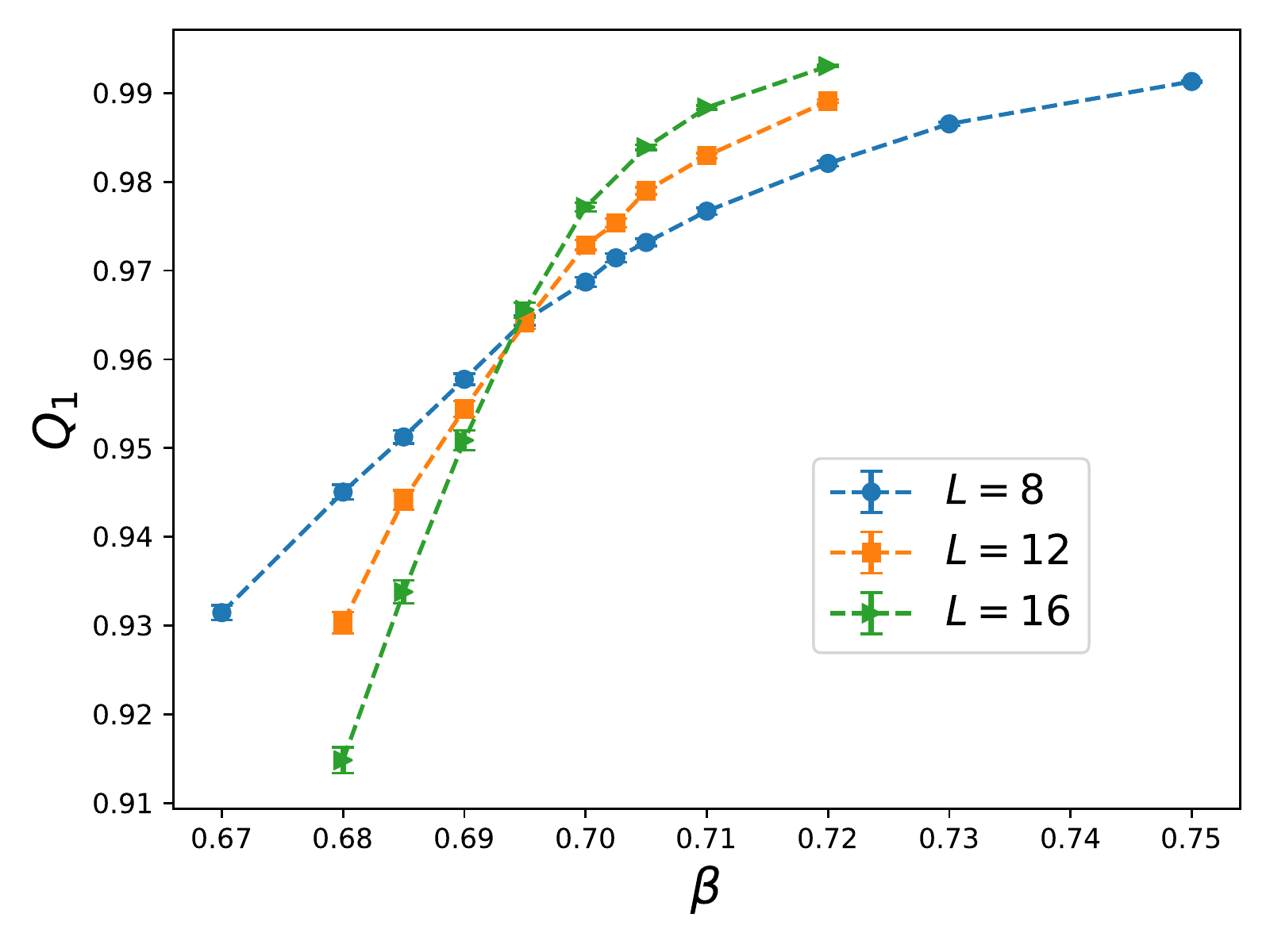}
\end{center}\vskip-0.7cm
\caption{$Q_1$ as functions of $\beta$ for the 3D classical $O(3)$ model.}
\label{O3_MC}
\end{figure}

In fig.~\ref{O3_MC}, the observable first Binder ratio $Q_1$
are considered as functions of $\beta$ for $L=8,12,16$. As can be seen from the figure, the curves corresponding to various $L$
intersect at a value of $\beta$ close to the predicted critical point $\beta_c = 0.6929$ \cite{Hol92,Cam02}.

For a $O(3)$ configuration obtained from the simulation, all the $S^3$ vectors associated with it are converted to $\psi$ mod $\pi$
and the resulting configuration is then fed into the trained NN.

\begin{figure}
\begin{center}
\vbox{~~~~~~~~~~~~~~~~~~~~
\includegraphics[width=0.5\textwidth]{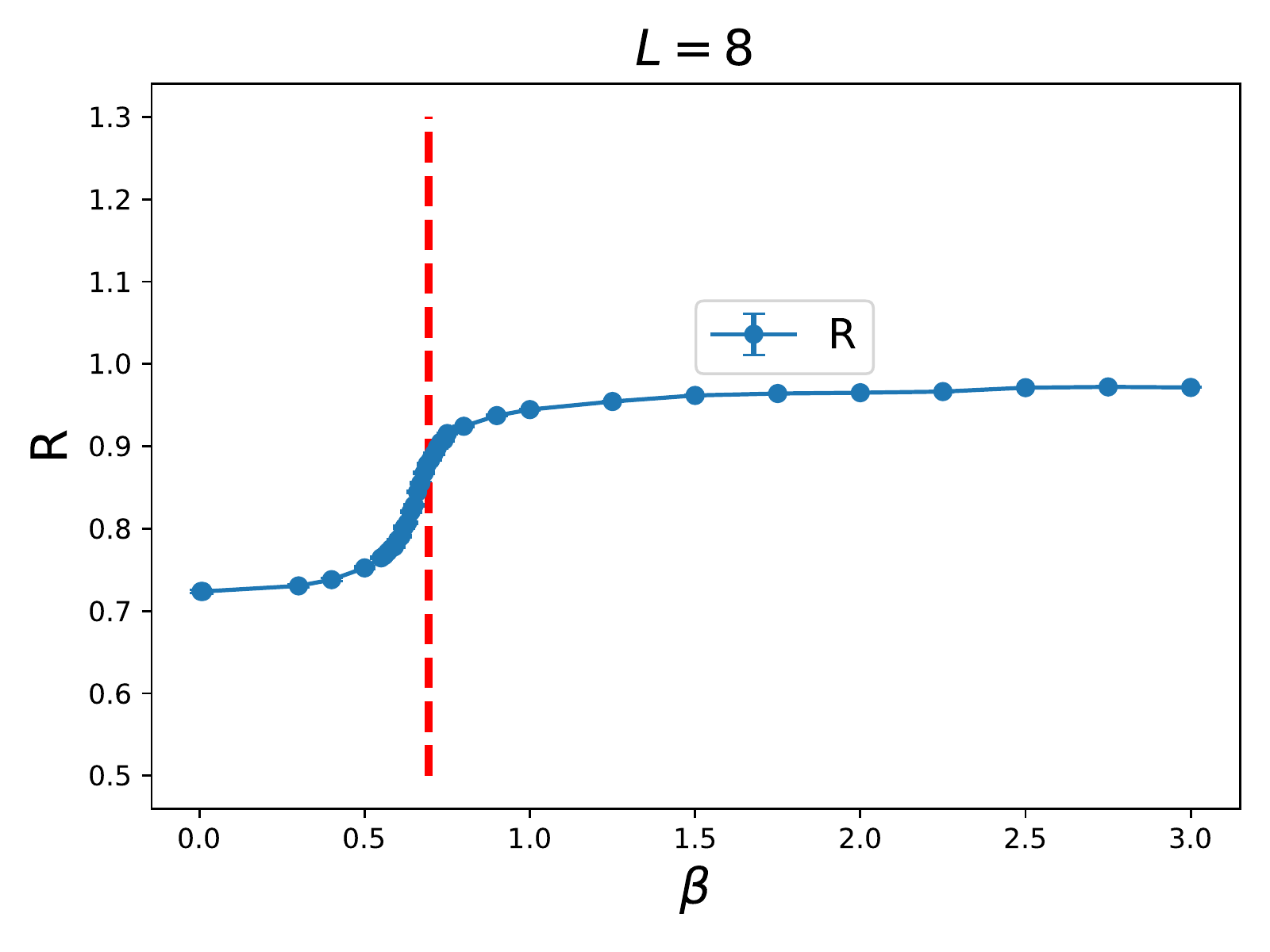}\vskip0.025cm
\includegraphics[width=0.5\textwidth]{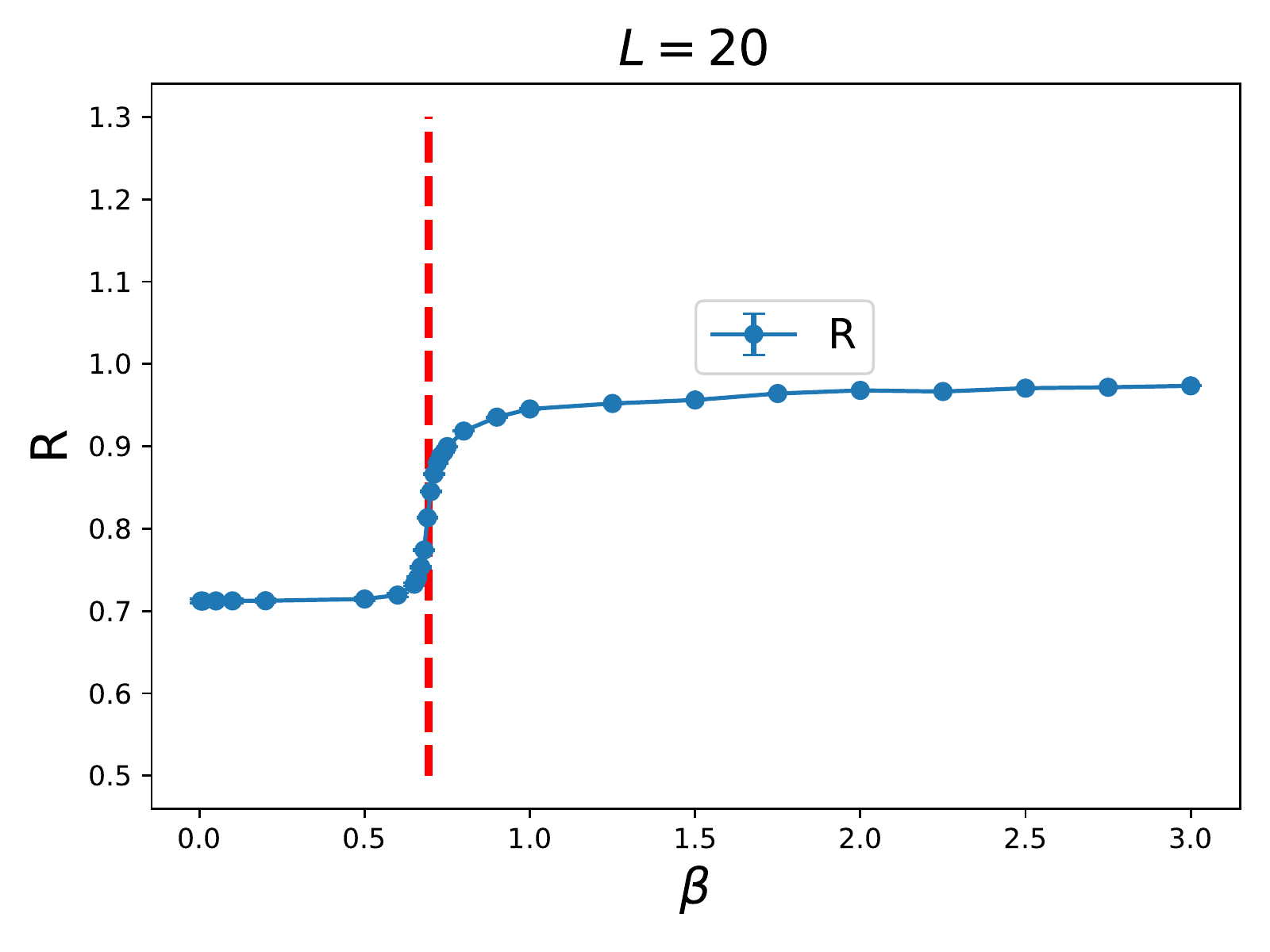}
}
\end{center}\vskip-0.7cm
\caption{$R$ as functions of $\beta$ for the 3D classical $O(3)$ model.
  The top and bottom panels are for $L=8$ and $L=20$, respectively.}
\label{Original_O3}
\end{figure}

$R$ as functions of $\beta$ for $L=8$ and $L=20$ are shown in fig.~\ref{Original_O3}. While it is clear that
both panels of fig.~\ref{Original_O3} imply $R$ change rapidly close to $\beta_c = 0.6929$,
$\beta_c$ cannot be calculated unambiguously when only the information of $R$ is available.

If one assumes that $R$ diminishes linearly with $\beta$ in the critical region, then $\beta_c$
can be approximately estimated by the intersection of the curves of $R$ and $1/\sqrt{2} + 1 - R$. Such an idea has been
used in Ref.~\cite{Tan20.1} to calculate the $T_c$ of the 3D 5-state ferromagnetic Potts model as well as the $g_c$ of the 3D plaquette model
(the latter will be studied in more detail here).
Here we adopt a more appropriate approach for the determination of the considered critical points by
taking into account the deviation between the theoretical and the calculated $R$.

\begin{figure}
\begin{center}
\vbox{
~~~~~~~~~~~~~~~~~
\includegraphics[width=0.5\textwidth]{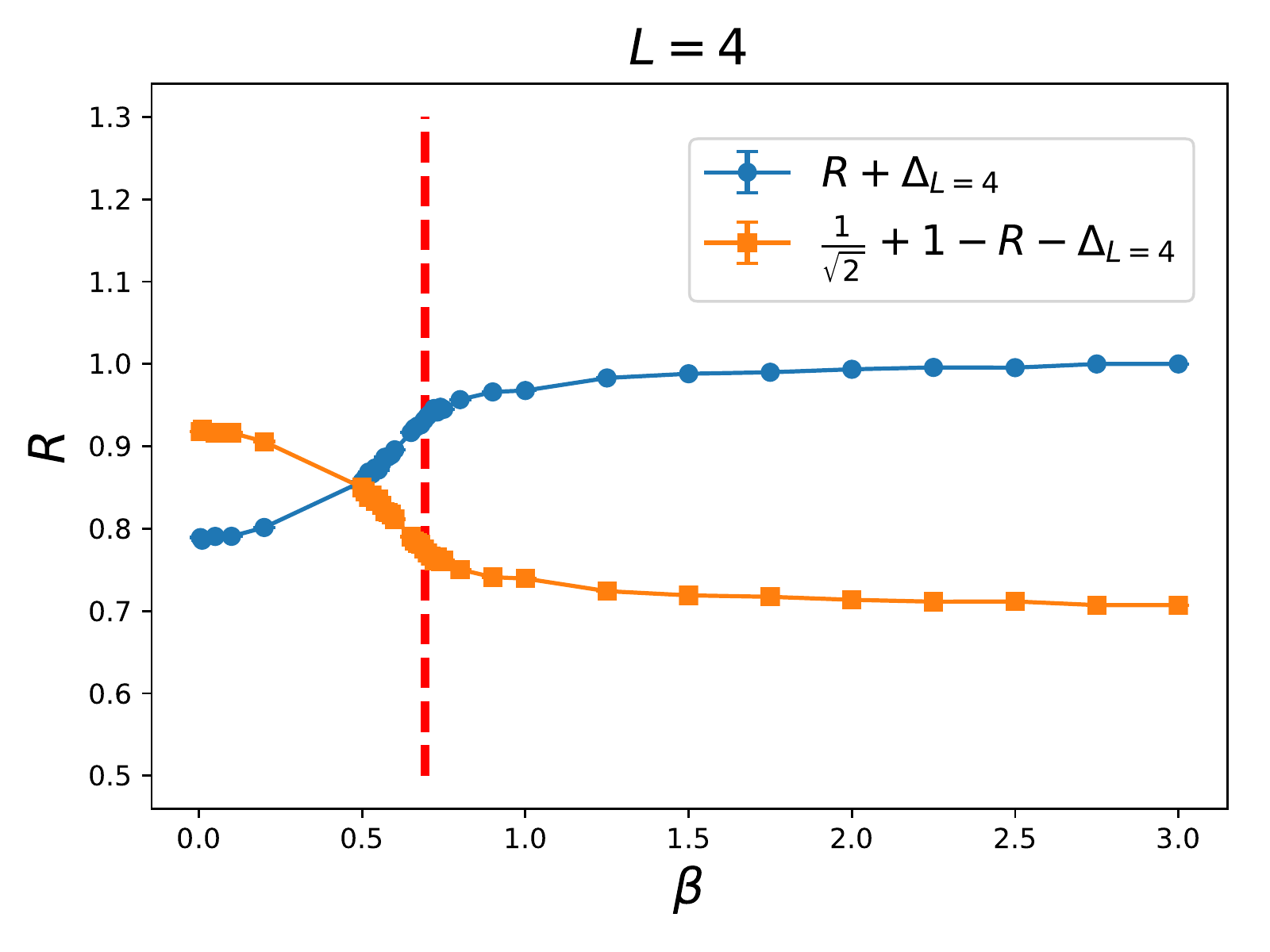}\vskip0.025cm
\includegraphics[width=0.5\textwidth]{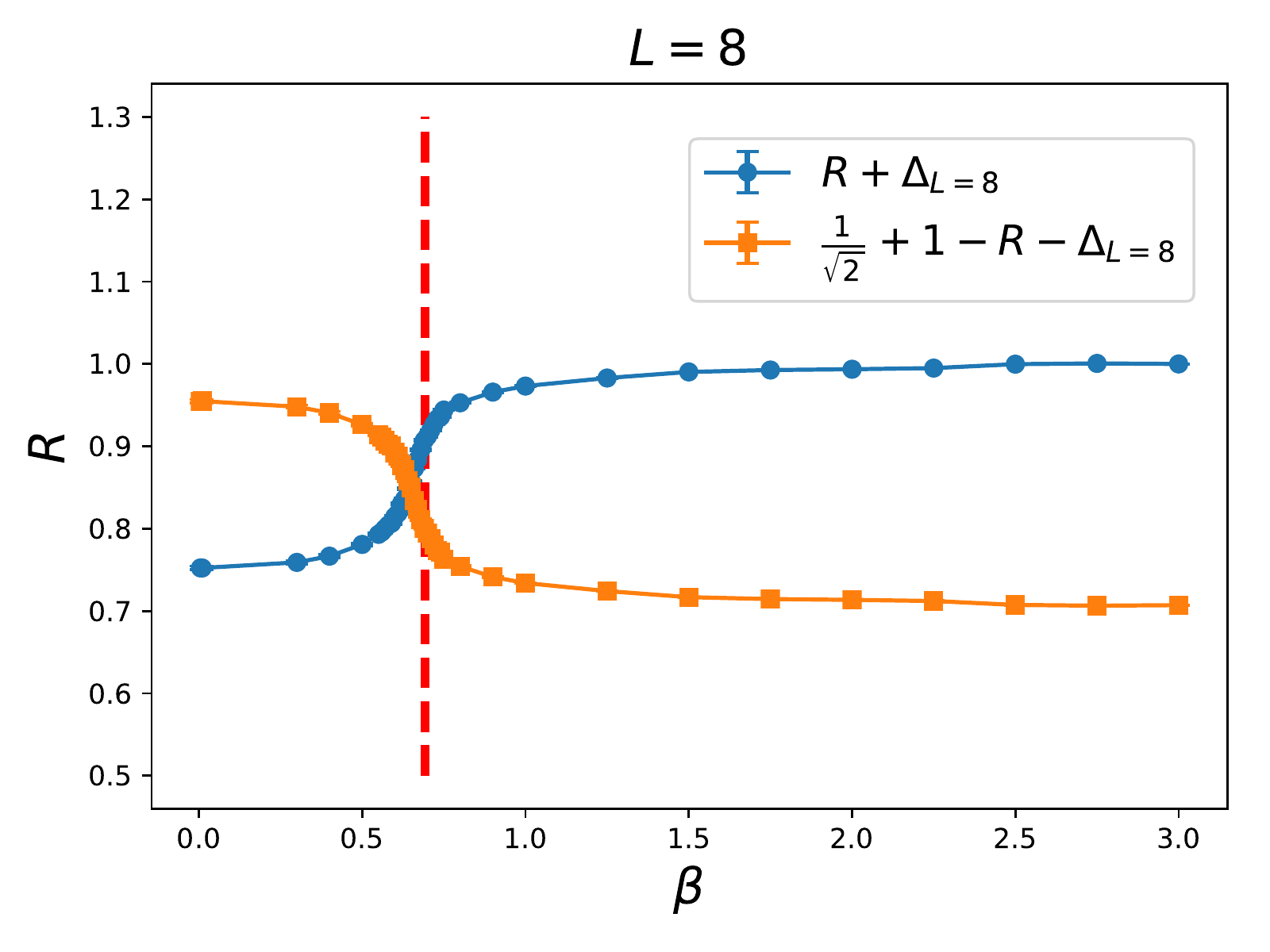}
}
\end{center}\vskip-0.7cm
\caption{$R+\Delta$ and $1/\sqrt{2}-\Delta + 1- R$ as functions of $\beta$ for
  the 3D classical $O(3)$ model. The top and bottom panels are for $L=4$ and $L=8$, respectively.}
\label{shifted_O3_1}
\end{figure}

\begin{figure}
\begin{center}
\vbox{
~~~~~~~~~~~~~~~~~
\includegraphics[width=0.5\textwidth]{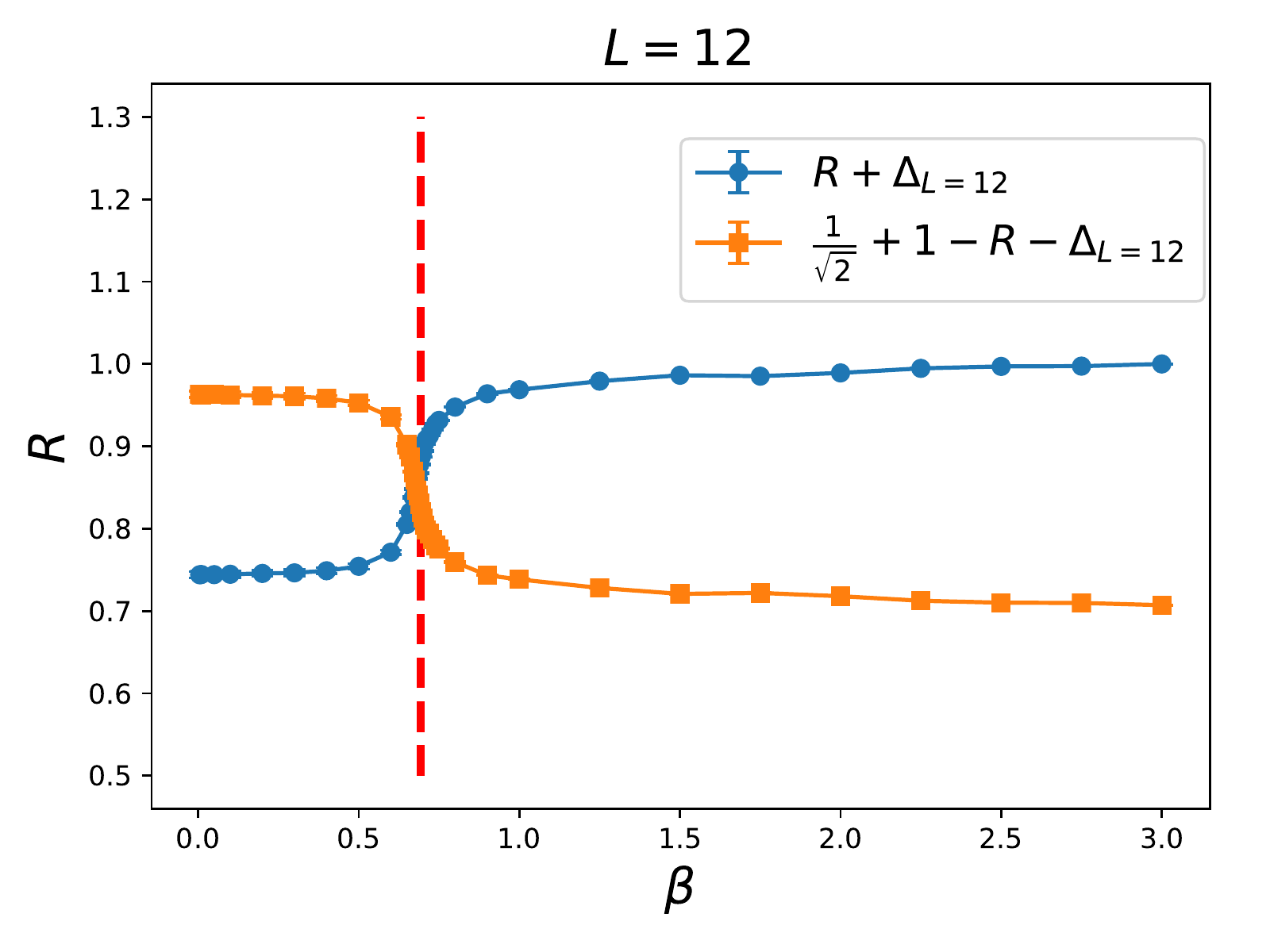}\vskip0.025cm
\includegraphics[width=0.5\textwidth]{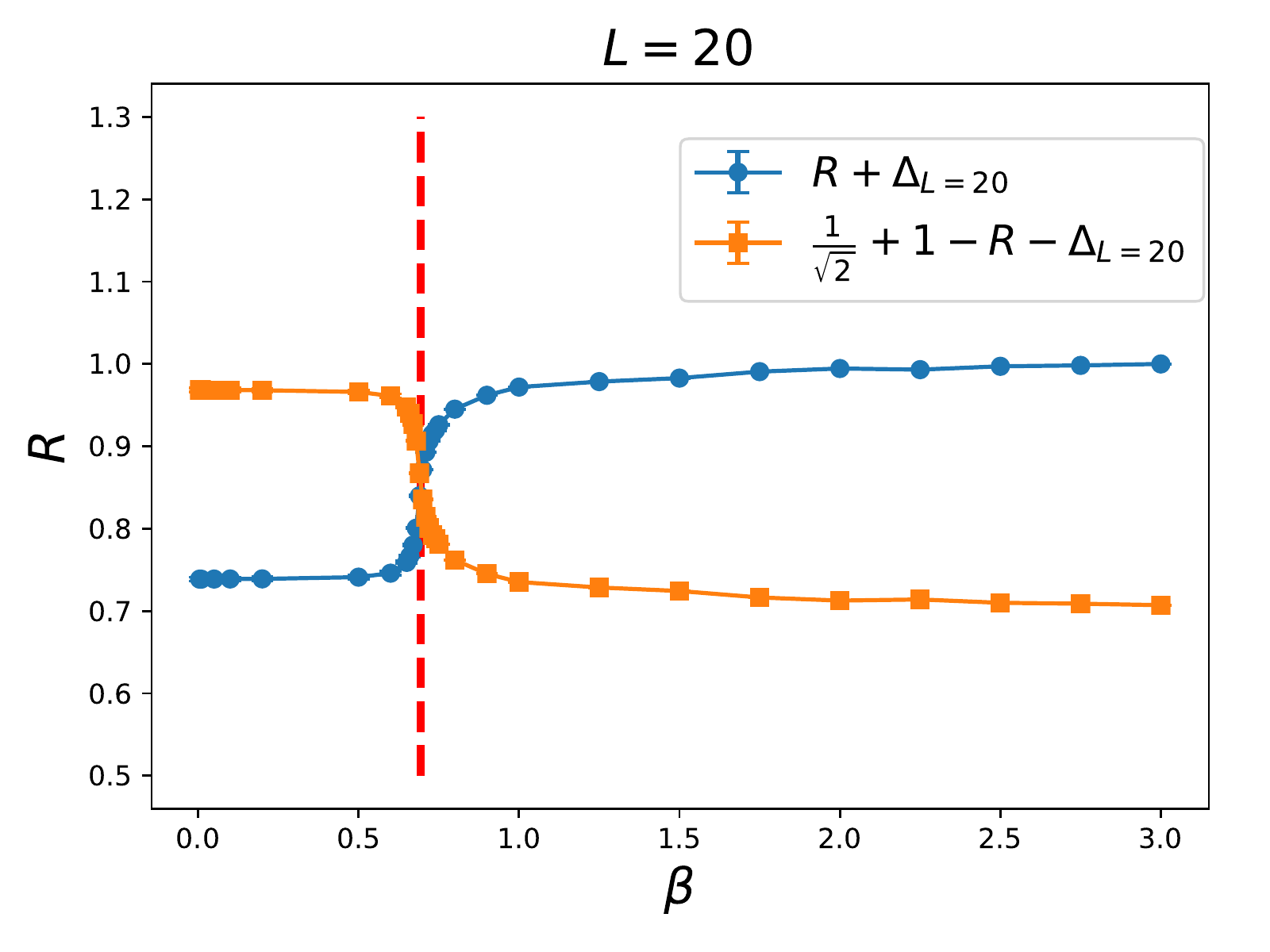}\vskip0.025cm
\includegraphics[width=0.5\textwidth]{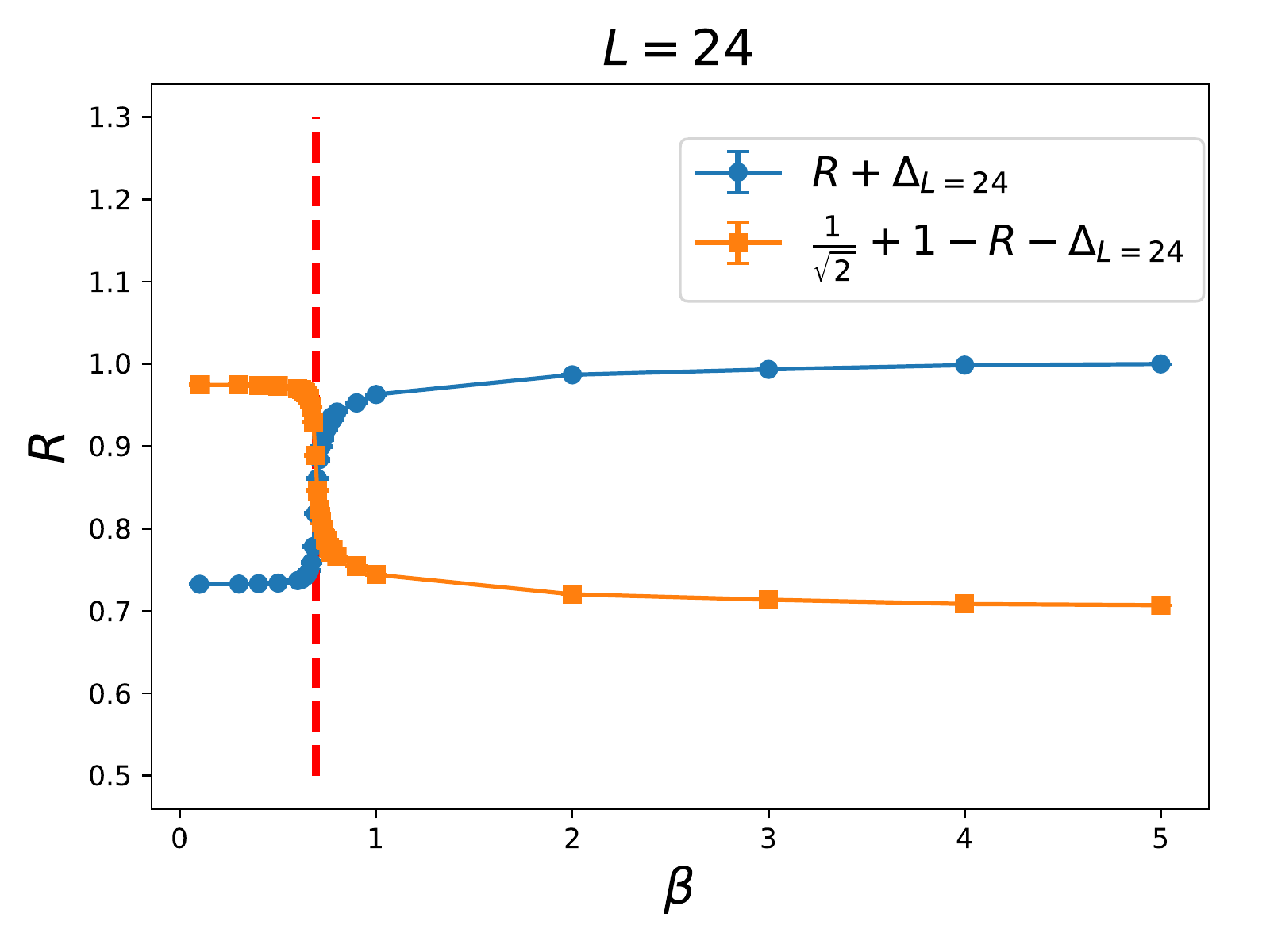}
}
\end{center}\vskip-0.7cm
\caption{$R+\Delta$ and $1/\sqrt{2}-\Delta + 1- R$ as functions of $\beta$ for
  the 3D classical $O(3)$ model. The top, middle, and bottom panels are for $L=12$, $L=20$, and $L=24$, respectively.}
\label{shifted_O3_2}
\end{figure}

Ideally, at extremely low temperature region, the obtained $R$ should be 1. To fulfill this criterion,
an overall shift $\Delta$, which is the difference between $1$ and the $R$
from the simulation with the largest $\beta$,
is conducted \cite{delta}. Figures~\ref{shifted_O3_1} and \ref{shifted_O3_2} demonstrate the
associated curves made up of considering the data
of $R+\Delta$ and $1/\sqrt{2} + 1 - R - \Delta$ as functions
of $\beta$ for $L = 4,8,12,20,24$. As can be seen from the figures, the intersections of these two curves for all the $L$ (except the one of $L=4$)
are in good
agreement with the theoretical prediction $\beta_c \sim 0.693$ (which are the vertical dashed lines in these figures).
While for large $L$, the estimated values of $\beta$ at which the
mentioned two curves intersect are slightly away from  $\beta_c = 0.6929$, the results shown in figs.~\ref{shifted_O3_1} and \ref{shifted_O3_2}
indicate that the idea of estimating $\beta_c$ by considering
the intersection of the curves associated with $R+\Delta$ and $1/\sqrt{2} + 1 - R - \Delta$ is an effective approach. In particular,
considering the simplicity of both the training procedure and the semi-experimental method of calculating $\beta_c$ (for any finite $L$)
employed in this study, the achievement reaches here for the determination of the $T_c$ of the complicated 3D classical $O(3)$ model is remarkable.

The success of calculating the $T_c$ of 3D classical $O(3)$ model through the idea of only considering $\psi$ mod $\pi$ indicates that
partial information of the model is sufficient to estimate its associated critical point accurately.

To calculate the critical points of the studied models with high precision using the intersections described above, 
one may apply certain forms of finite-size scaling to those crossing points. Based on the outcomes demonstrated in
figs.~\ref{shifted_O3_1} and \ref{shifted_O3_2}, it is clear that the $R$ associated with the 3D classical $O(3)$ model receives
mild finite-size effect. Apart from this, accurate determination
of the crossing places in the relevant parameter space, particularly high precision estimated uncertainties for these crossing points,
is needed in order to carry out the fits.
Hence we postpone such an analysis to a latter subsection where 2D 3-state
ferromagnetic Potts model and 2D classical XY models on the square lattices
are discussed.

\begin{figure}
\begin{center}
\includegraphics[width=0.5\textwidth]{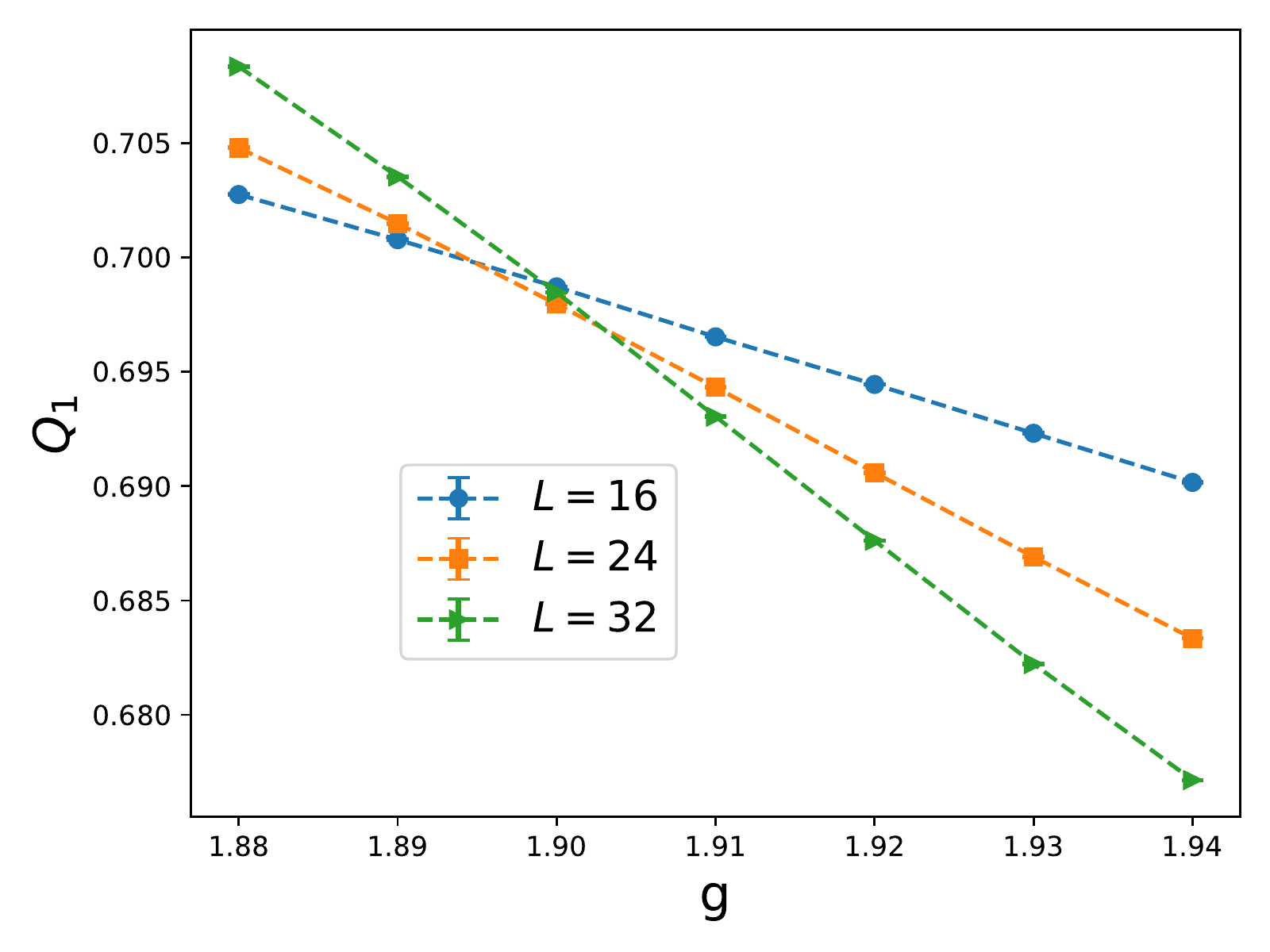}
\end{center}\vskip-0.7cm
\caption{$Q_1$ (of $L=16,24,32$) as functions of $g$ for the 2D dimerized
  quantum ladder model.}
\label{ladder_MC}
\end{figure}

\subsection{Results of 2D quantum spin system}

The first Binder ratio $Q_1$ close to $g_c$ for the studied 2D dimerized spin-1/2 antiferromagnet (2D ladder model) are shown in fig.~\ref{ladder_MC}.
Similar to the case of
3D classical $O(3)$ model, various curves of large $L$ tend to intersect at a value of $g$ around 1.9. The estimated intersection $g \sim$ 1.9
matches nicely with the known result $g_c = 1.90948(5)$ in the literature \cite{San10}. Of course, a better determination of $g_c$ requires
the performance of a dedicated finite-size scaling analysis. 

\begin{figure}
\begin{center}
\vbox{
~~~~~~~~~~~~~~~~~
\includegraphics[width=0.4\textwidth]{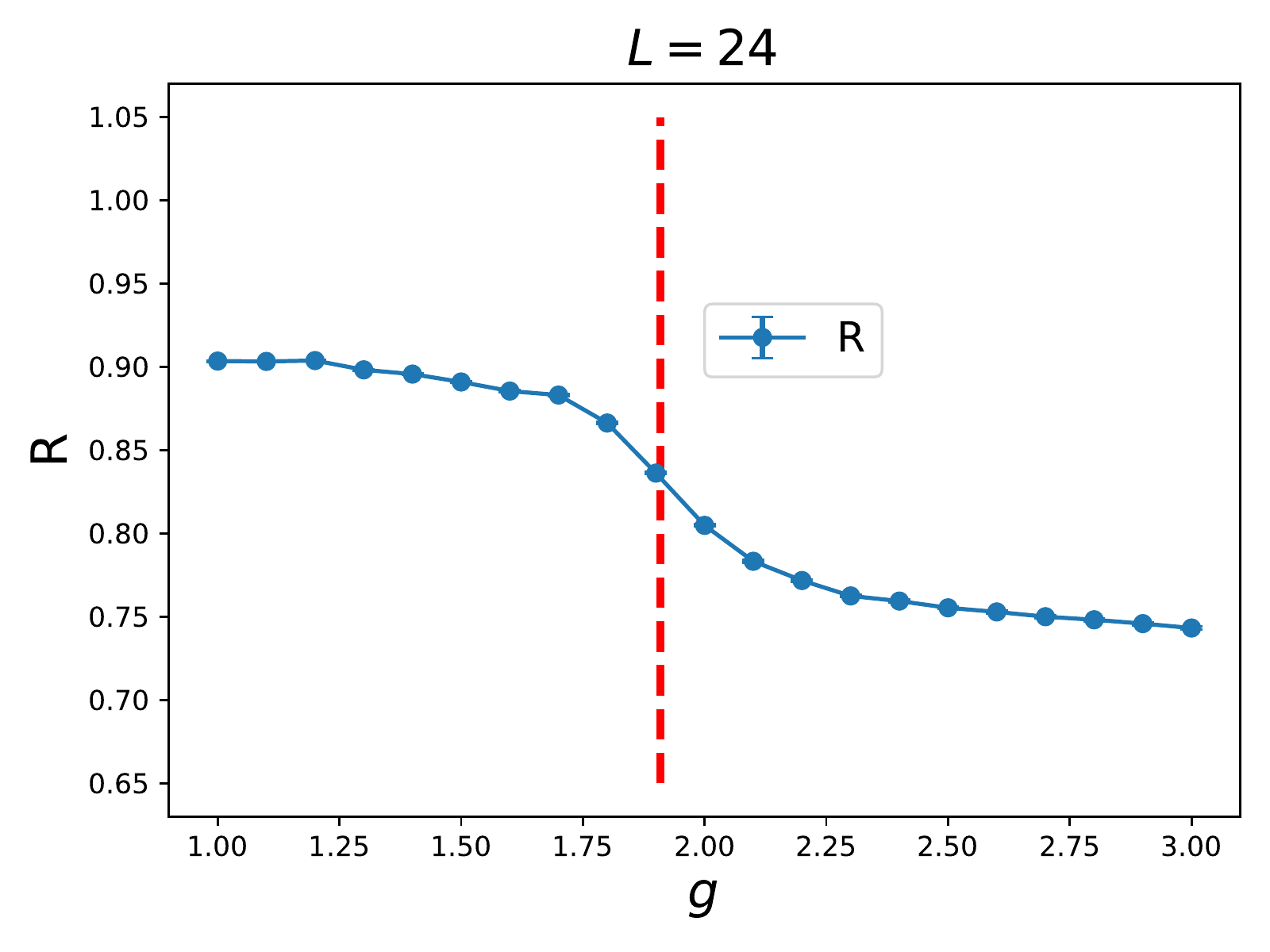}\vskip0.025cm
\includegraphics[width=0.4\textwidth]{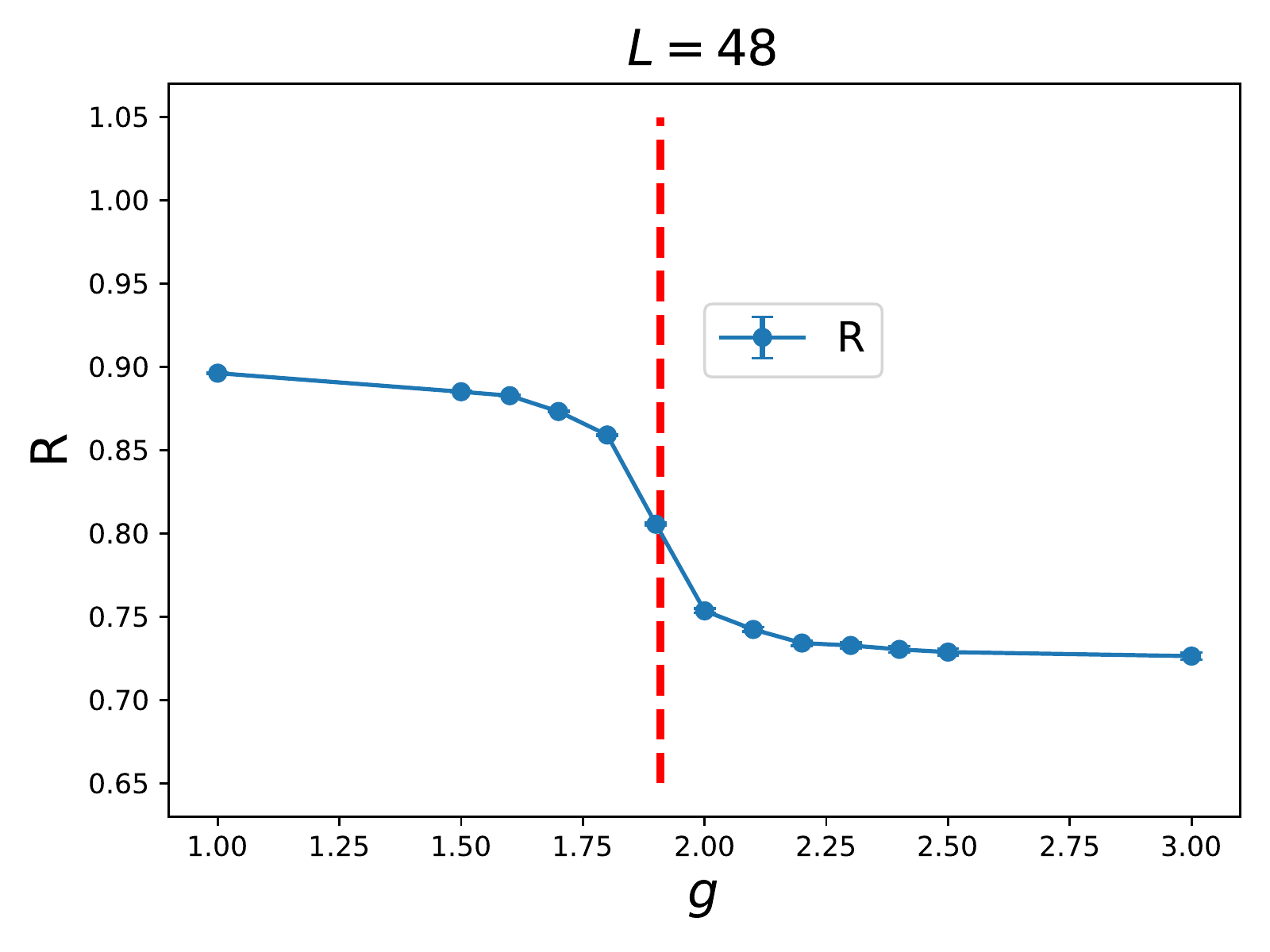}
}
\end{center}\vskip-0.7cm
\caption{$R$ as functions of $g$ for the 2D dimerized quantum ladder model. The top and bottom panels are for $L=24$ and $L=48$, respectively.}
\label{Original_ladder}
\end{figure}

\begin{figure}
\begin{center}
\vbox{
~~~~~~~~~~~~~~~~~
\includegraphics[width=0.5\textwidth]{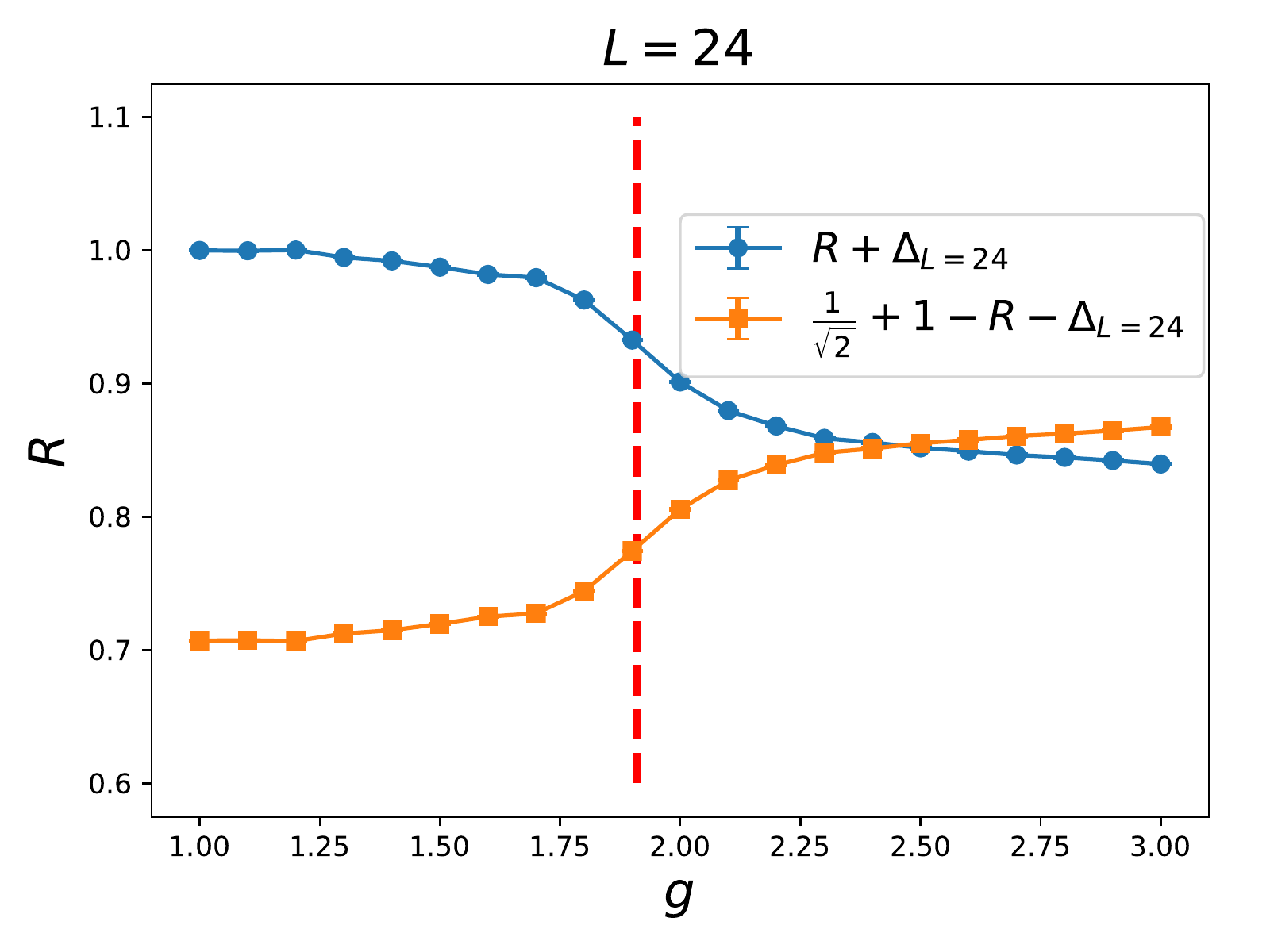}\vskip0.025cm
\includegraphics[width=0.5\textwidth]{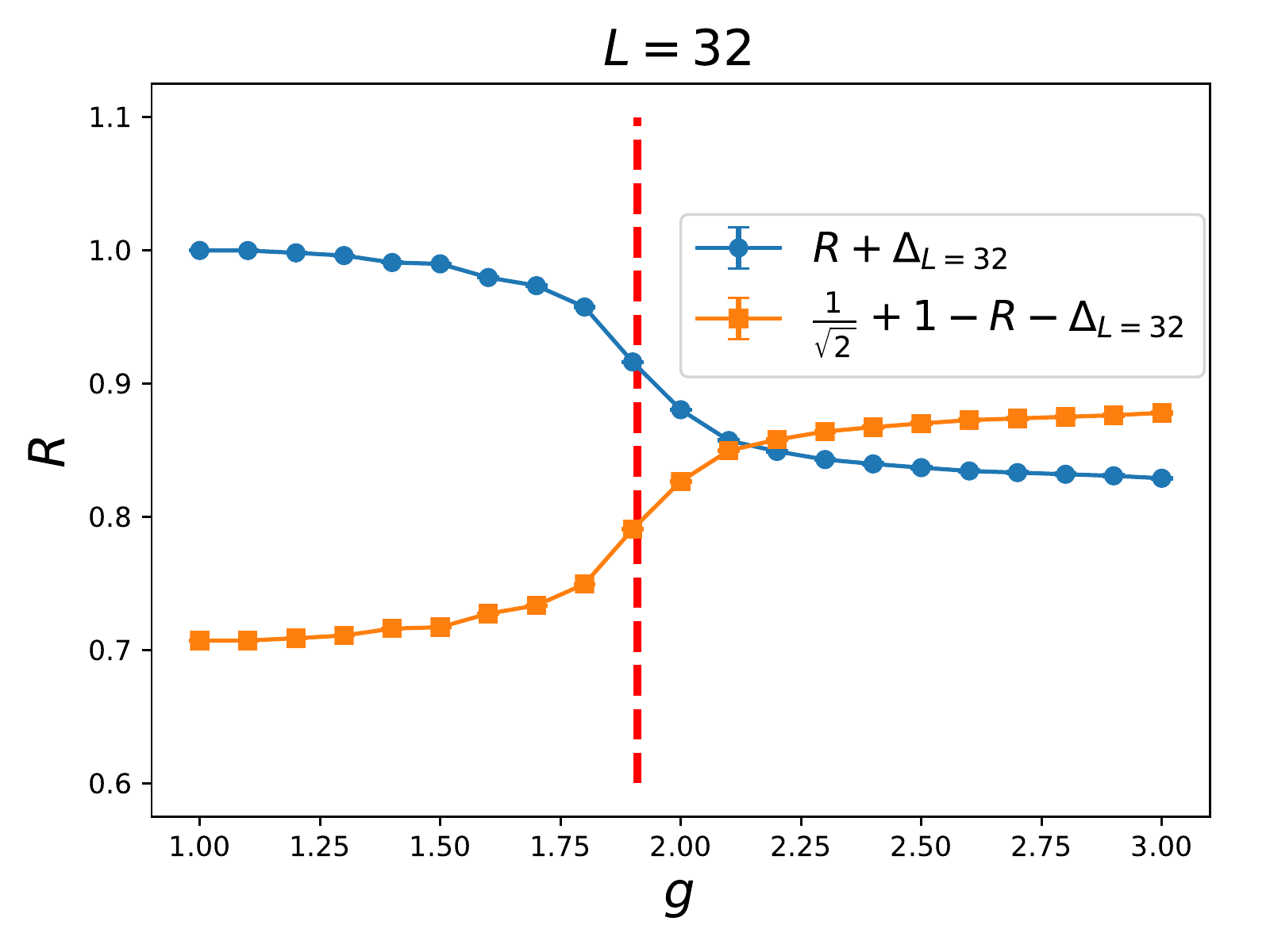}
}
\end{center}\vskip-0.7cm
\caption{$R+\Delta$ and $1/\sqrt{2}-\Delta + 1- R$ as functions of $g$ for
  the 2D dimerized quantum ladder model. The top and bottom panels are for $L=24$ and $L=32$, respectively.}
\label{shifted_ladder_1}
\end{figure}

\begin{figure}
\begin{center}
\vbox{
~~~~~~~~~~~~~~~~~
\includegraphics[width=0.5\textwidth]{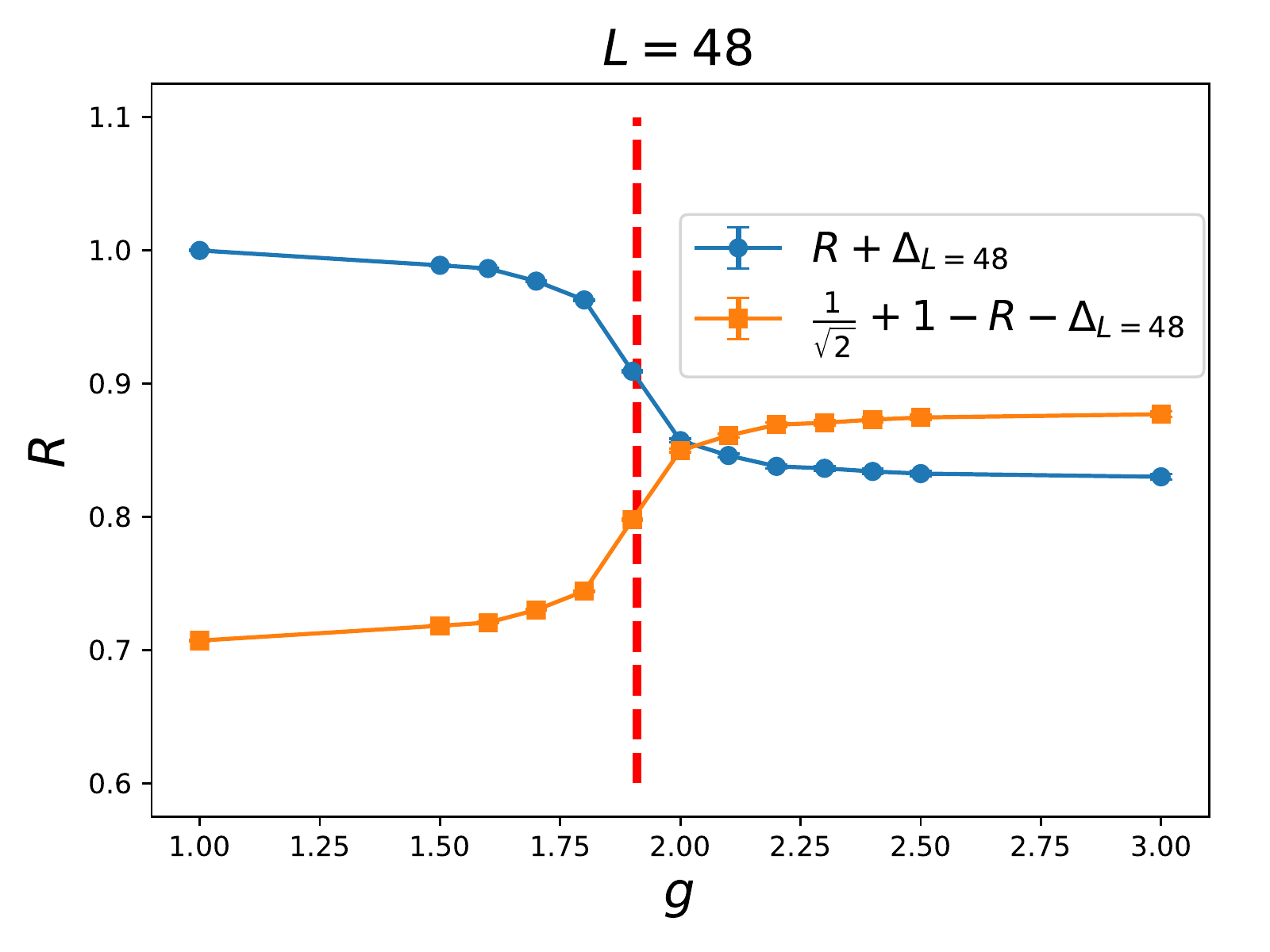}\vskip0.025cm
\includegraphics[width=0.5\textwidth]{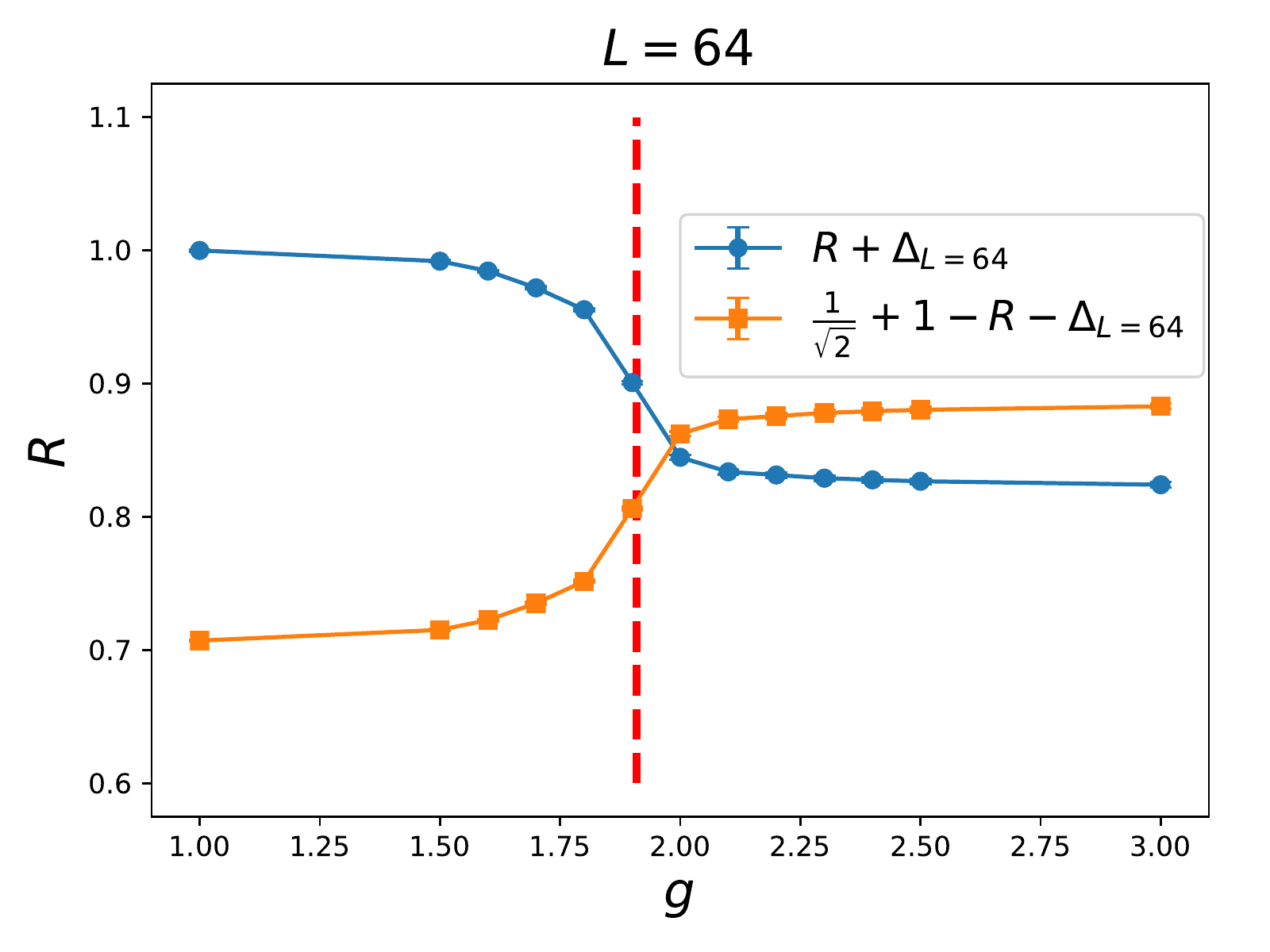}~~~~~~
}
\end{center}\vskip-0.7cm
\caption{$R+\Delta$ and $1/\sqrt{2}-\Delta + 1- R$ as functions of $g$ for
  the 2D dimerized quantum ladder model. The top and bottom panels are for $L=48$ and $L=64$, respectively.}
\label{shifted_ladder_2}
\end{figure}

For the 2D ladder model, the associated $R$ as functions of $g$ for $L=24,48$ are shown in fig.~\ref{Original_ladder}.
Moreover, by using the idea of estimating $\beta_c$ for the 3D classical $O(3)$ model,
the curves resulting from treating $R+\Delta$ and $1/\sqrt{2} + 1 - R - \Delta$ as functions of $g$
are demonstrated in
figs.~\ref{shifted_ladder_1} ($L=24,32$) and \ref{shifted_ladder_2} ($L=48,64$).
The vertical dashed lines in these figures are the theoretical $g_c$. Here $\Delta$ is the difference between
the theoretical and the calculated values of $R$ at $g=1$. As can be seen from the figures,
when box size $L$ increases, the $g$ at which the mentioned two curves intersects is approaching the theoretical $g_c$.
Hence the outcomes demonstrated in the figures support the fact that our method of determining the critical points is also valid for the
investigated quantum (spin) system.

\subsection{Results of 3D quantum spin system}

\begin{figure}
  \begin{center}
      \includegraphics[width=0.5\textwidth]{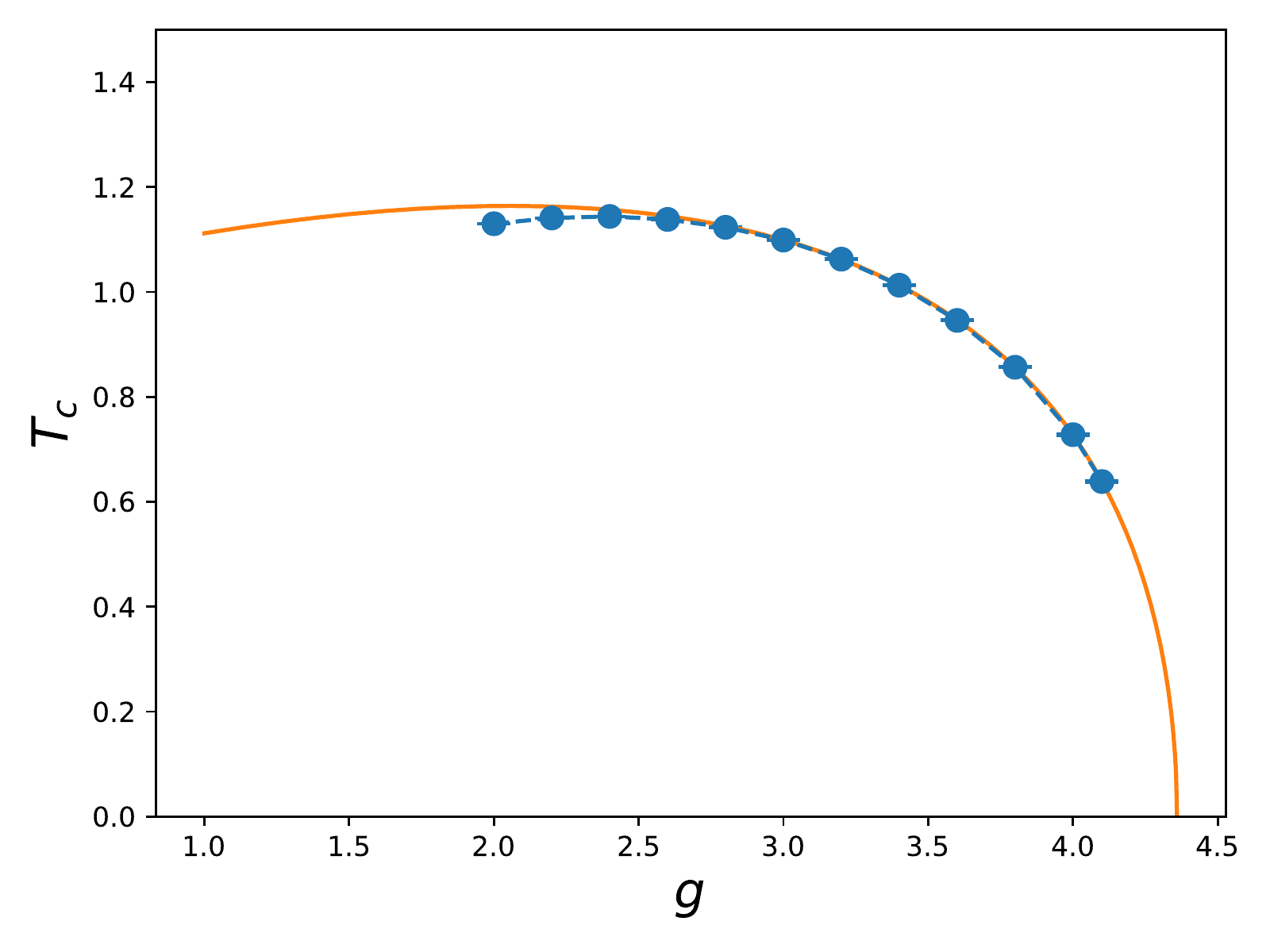}
\end{center}\vskip-0.7cm
\caption{$T_N$ as a function of $g$ for the 3D dimerized quantum plaquette model. 
The solid line shown in the figure is obtained by using the results from a fit.}
\label{3D_plaq_Jc}
\end{figure}

The $g_c$ of the 3D plaquette model studied in Ref.~\cite{Tan18} can be determined by considering the N\'eel temperatures $T_N$ of various $g$ close to $g_c$.
Specifically, if the logarithmic correction is not taken into account, then close to $g_c$, $T_N$ can be described by
$T_N \sim A|g-g_c|^{c}+B|g-g_c|^{2c}$, here $A$, $B$, and $c$ are some constants.
As a result, $g_c$ can be calculated by fitting the data of $T_N$ of various $g$ to this form. The $g_c$ estimated by this approach lies
between 4.35 and 4.375, see fig.~\ref{3D_plaq_Jc}. This obtained $g_c$ will be used to examine the effectiveness of the NN method of calculating the $g_c$ of the 3D
plaquette model.

\begin{figure}
\begin{center}
\vbox{
~~~~~~~~~~~~~~~~~
\includegraphics[width=0.5\textwidth]{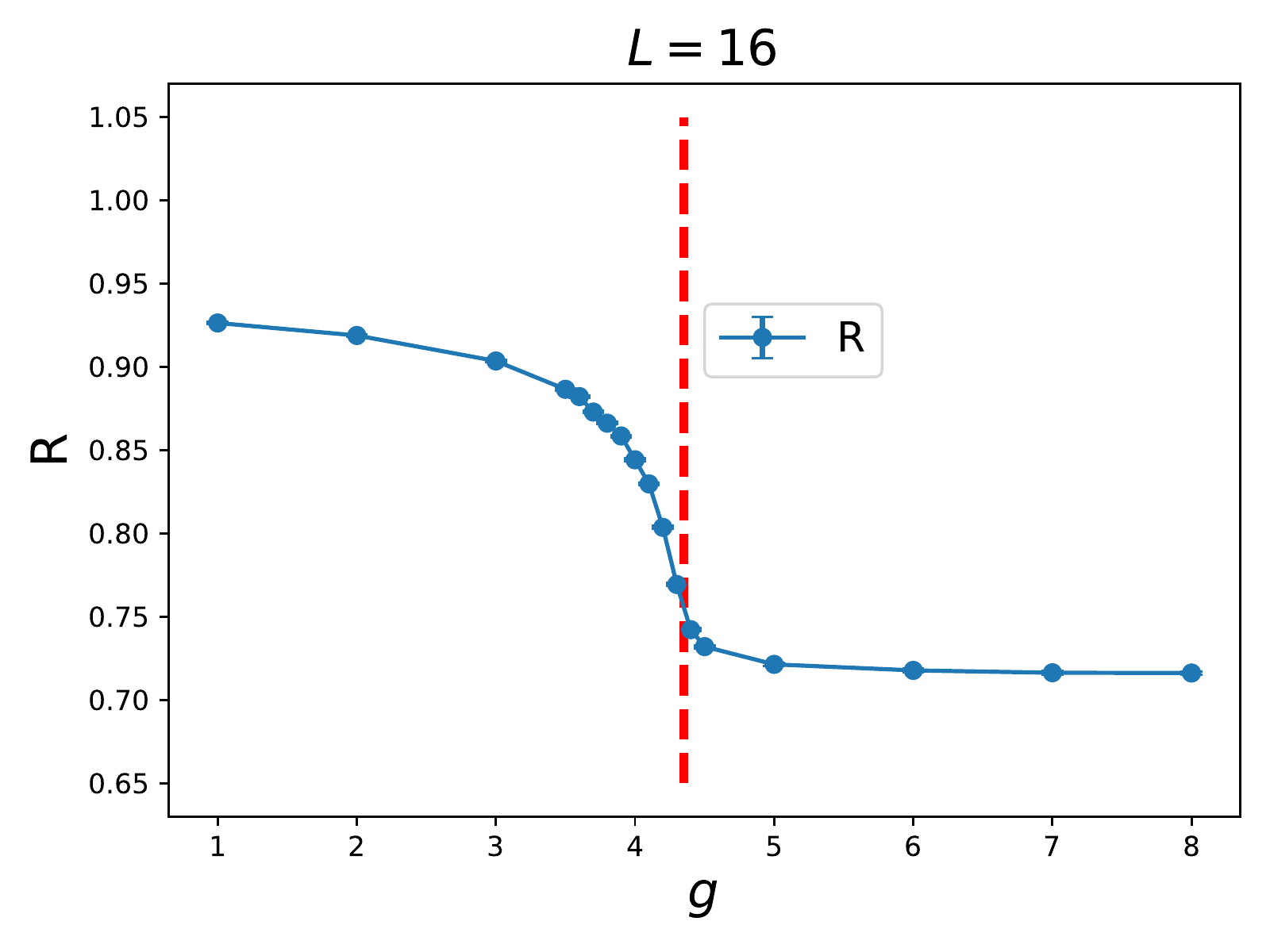}\vskip0.025cm
\includegraphics[width=0.5\textwidth]{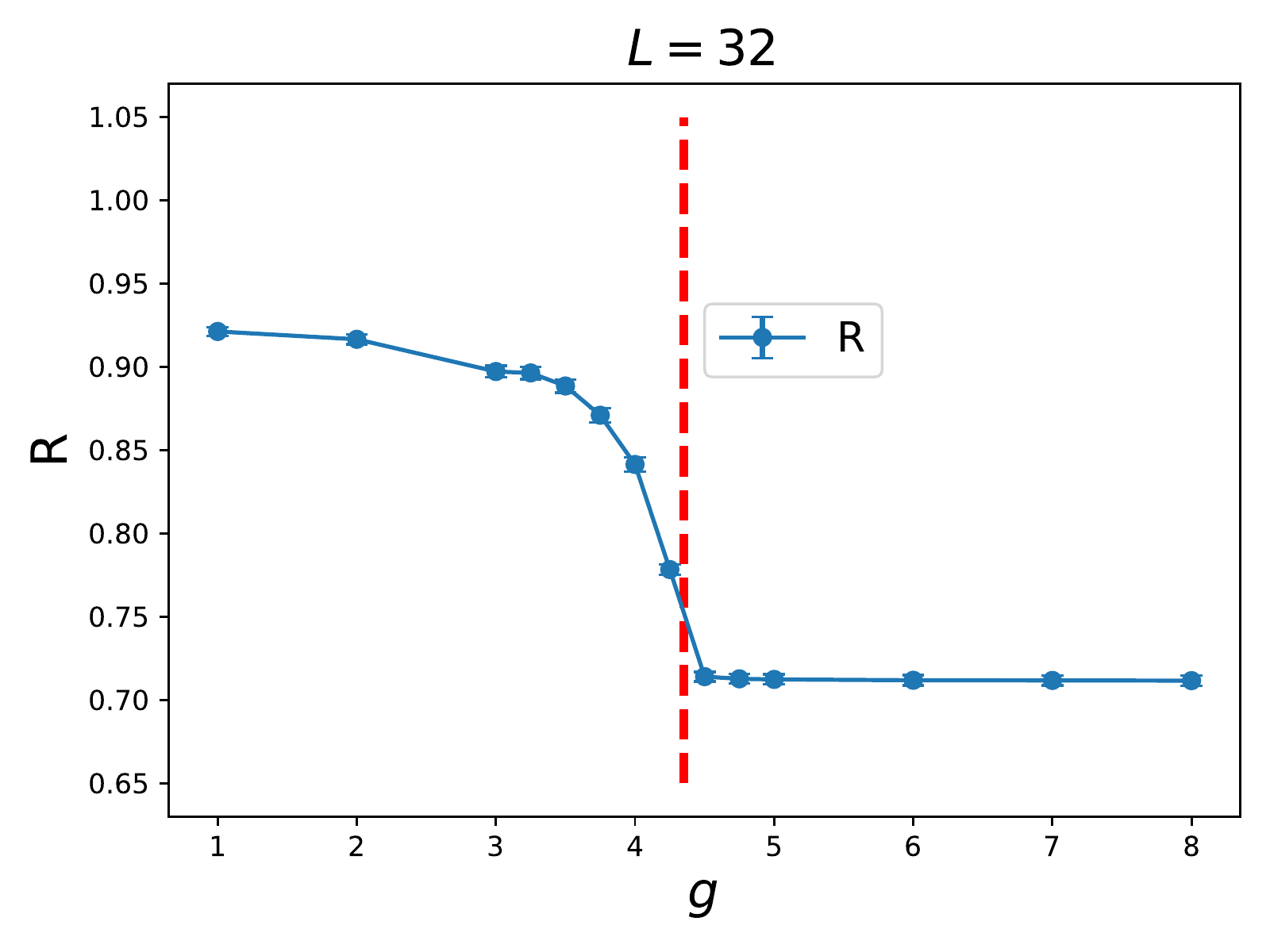}
}
\end{center}\vskip-0.7cm
\caption{$R$ as functions of $g$ for the 3D dimerized quantum plaqutte model. The top and bottom panels are for $L=16$ and $L=32$, respectively.}
\label{Original_plaquette}
\end{figure}

\begin{figure}
\begin{center}
\vbox{
~~~~~~~~~~~~~~~~~
\includegraphics[width=0.5\textwidth]{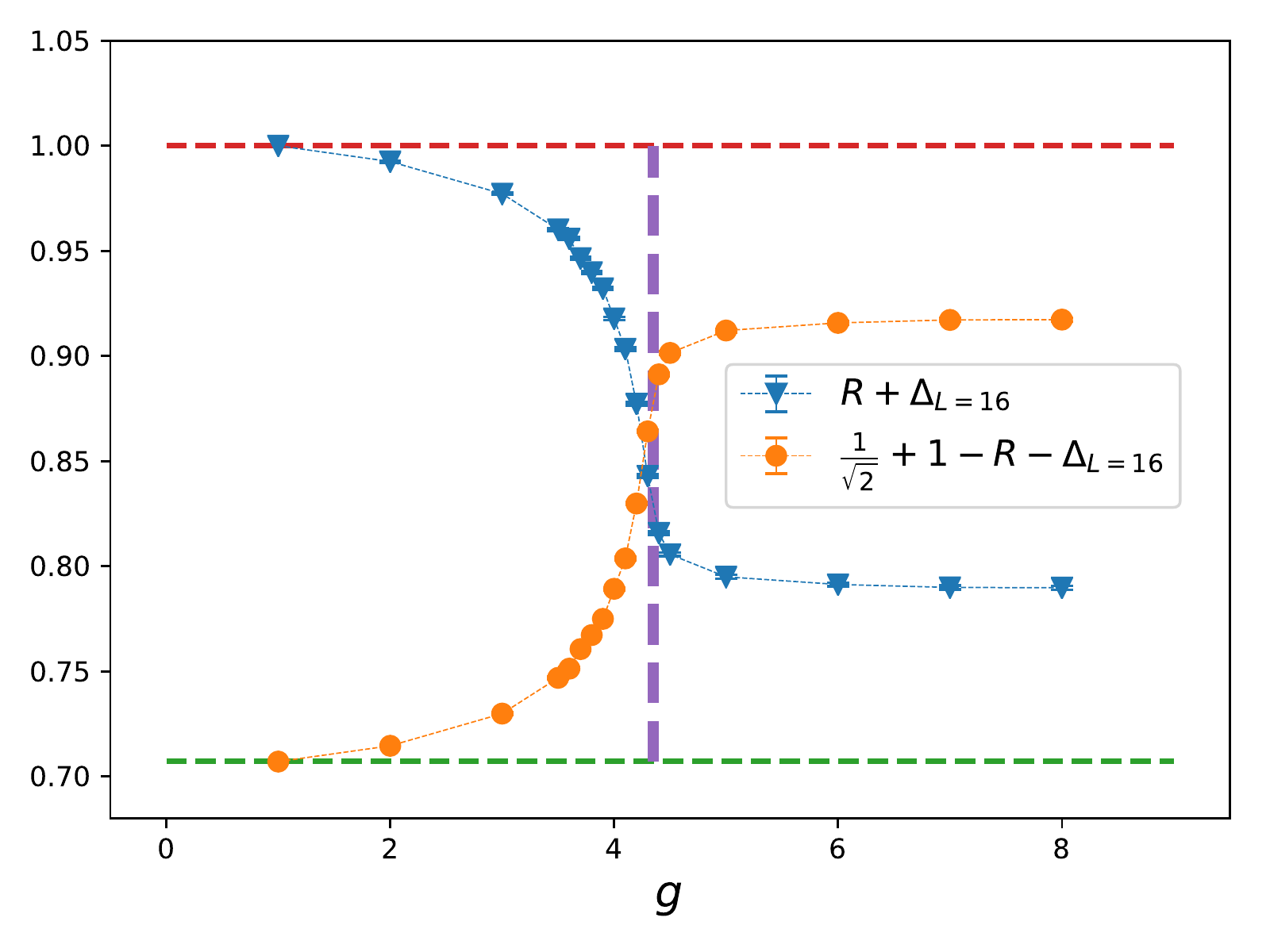}\vskip0.025cm
\includegraphics[width=0.5\textwidth]{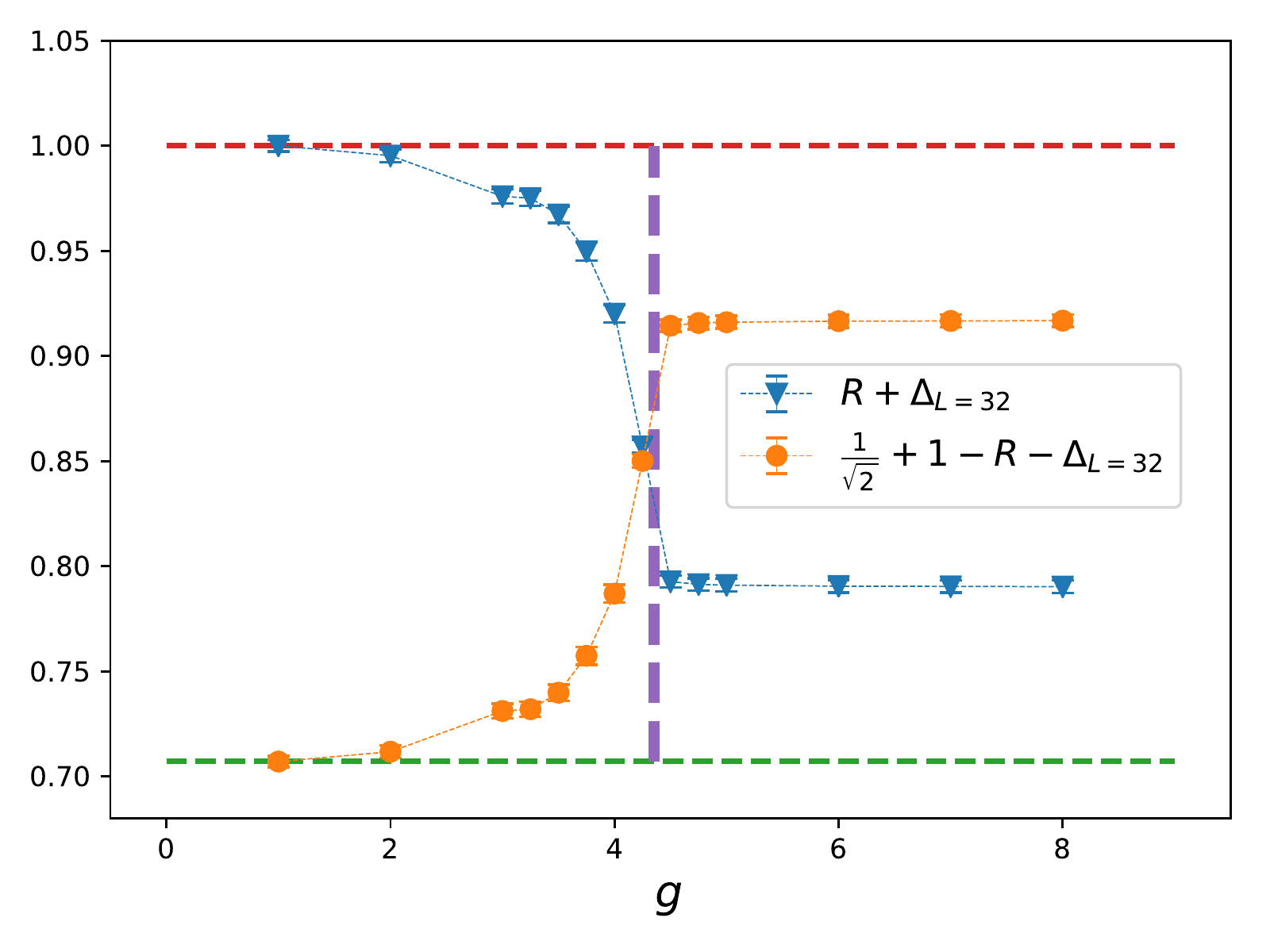}~~~~~~
}
\end{center}\vskip-0.7cm
\caption{$R+\Delta$ and $1/\sqrt{2}-\Delta + 1- R$ as functions of $g$ for
  the 3D dimerized quantum plaquette model. The top and bottom panels are for $L=16$ and $L=32$, respectively.}
\label{shifted_plaquette_1}
\end{figure}

$R$ as functions of $g$ for $L=16$ and 32 for the 3D plaquette model are shown in fig.~\ref{Original_plaquette}. In addition, the curves resulting from considering
$R+\Delta$ and
$1/\sqrt{2} + 1 - R - \Delta$ as functions of $g$ are demonstrated in fig.~\ref{shifted_plaquette_1} ($L=16,32$). The vertical dashed lines
in these figures are 4.35 which is the estimated lower bound for $g_c$ discussed in the previous paragraph. Here $\Delta$ is again
the difference between the theoretical and the calculated values of $R$ at $g=1$.

Remarkably, just like what we have found for the 2D quantum ladder model, the results shown in the figure clearly reveal the message that
our NN method is valid for 3D quantum spin model as well.

It is interesting to notice that the crossing points in both panels of fig.~\ref{shifted_plaquette_1} are slightly below the critical
point calculated from $T_N$. We attribute this to the facts that the systematic influence of some tunable parameters of NN as well as
certain corrections to the employed finite-size scaling method are not taken into account here.  

Nevertheless, based on the outcomes associated with both the investigated 2D and 3D dimerized quantum antiferromagnetic Heisenberg models,
it is beyond doubt that the NN approach employed here can be used to estimate the critical points of quantum phase transitions efficiently.

\subsection{Verification of the semi-experimental finite-size scaling formulas: 2D three-state
  ferromagnetic Potts model and 2D classical XY model on the square lattices}

\subsubsection{2D three-state ferromagnetic Potts model}

In previous subsections, it is shown that the critical point can be obtained by considering
the intersection of two curves made up of quantities associated with $R$. To obtain a high
precision estimation for the critical point using the crossing points, one can apply
certain expression of finite-size scaling to fit the data (of the crossing points). Here
we use the data of the 2D 3-state ferromagnetic Potts model on the square lattice available
in Ref.~\cite{Li18} to carry out such an investigation. For each $L$, the data are obtained using
a single set of relevant NN parameters. As a result, the quoted errors are associated
with the Potts configurations themselves.

\begin{figure}
\begin{center}
\includegraphics[width=0.5\textwidth]{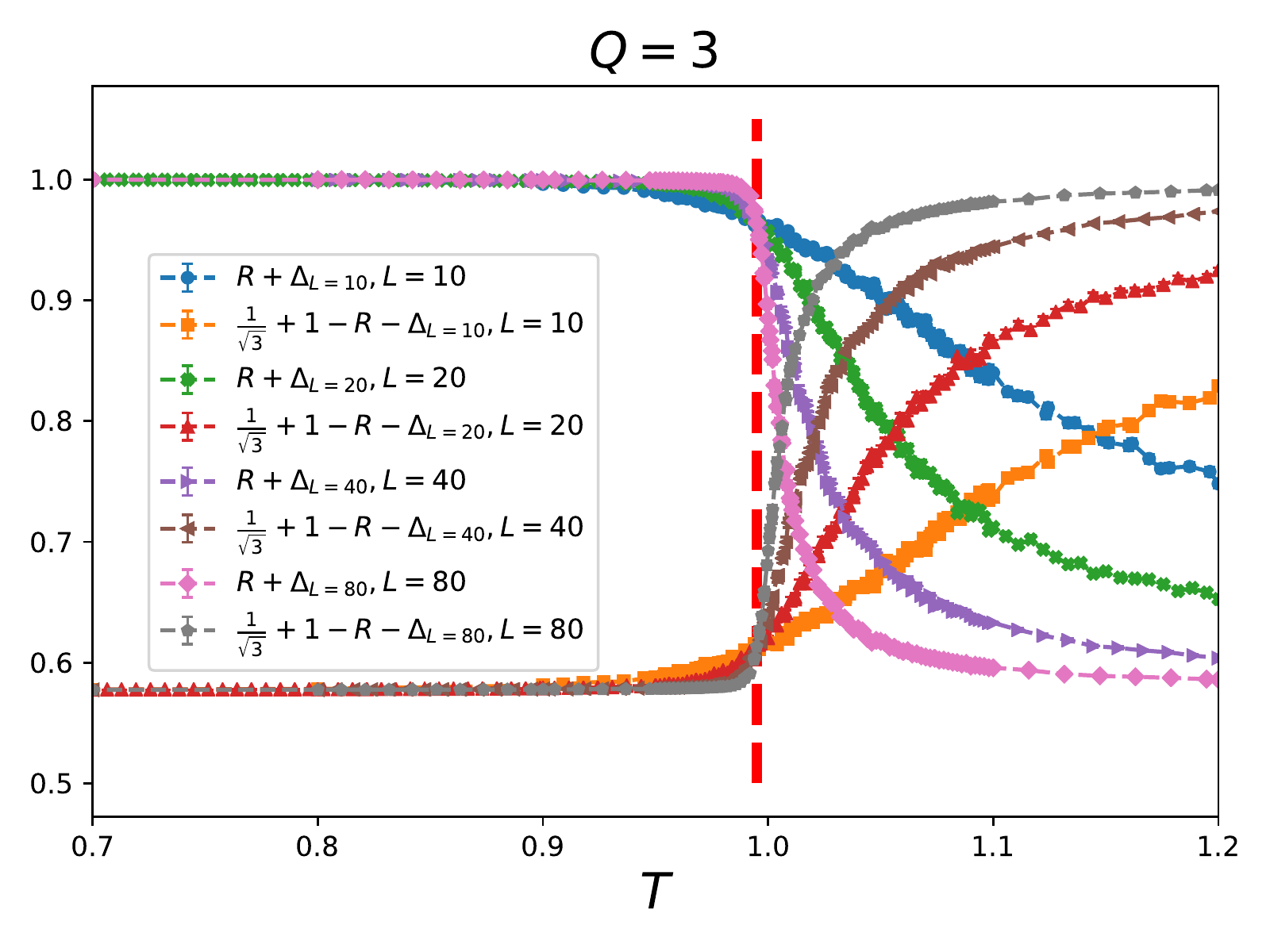}
\end{center}\vskip-0.7cm
\caption{$R+\Delta$ and $1/\sqrt{3} + 1 - R -\Delta$ as functions of $T$ for various $L$
  for the 2D 3-state ferromagnetic Potts model on the square lattice.}
\label{Q3R}
\end{figure}

Figure~\ref{Q3R} shows that data of $R+\Delta$ and $1/\sqrt{3} + 1 - R -\Delta$ as
functions of $T$ for various $L$ for the 2D 3-state ferromagnetic Potts model
on the square lattice \cite{Li18}. A fit of the form $a + b/L^{c} $, where
$a$, $b$, and $c$ are some to be determined constants ($a$ is exactly the desired $T_c$), is used to fit the data
of the crossing points obtained from $L=10,20,40,80,120,240$ (The data
of $L=120$ and 240 are not presented in fig.~\ref{Q3R}). When carrying out the fits, Gaussian noises are considered
in order to estimate the corresponding errors of the constants $a$, $b$, and $c$.

\begin{figure}
\begin{center}
\includegraphics[width=0.5\textwidth]{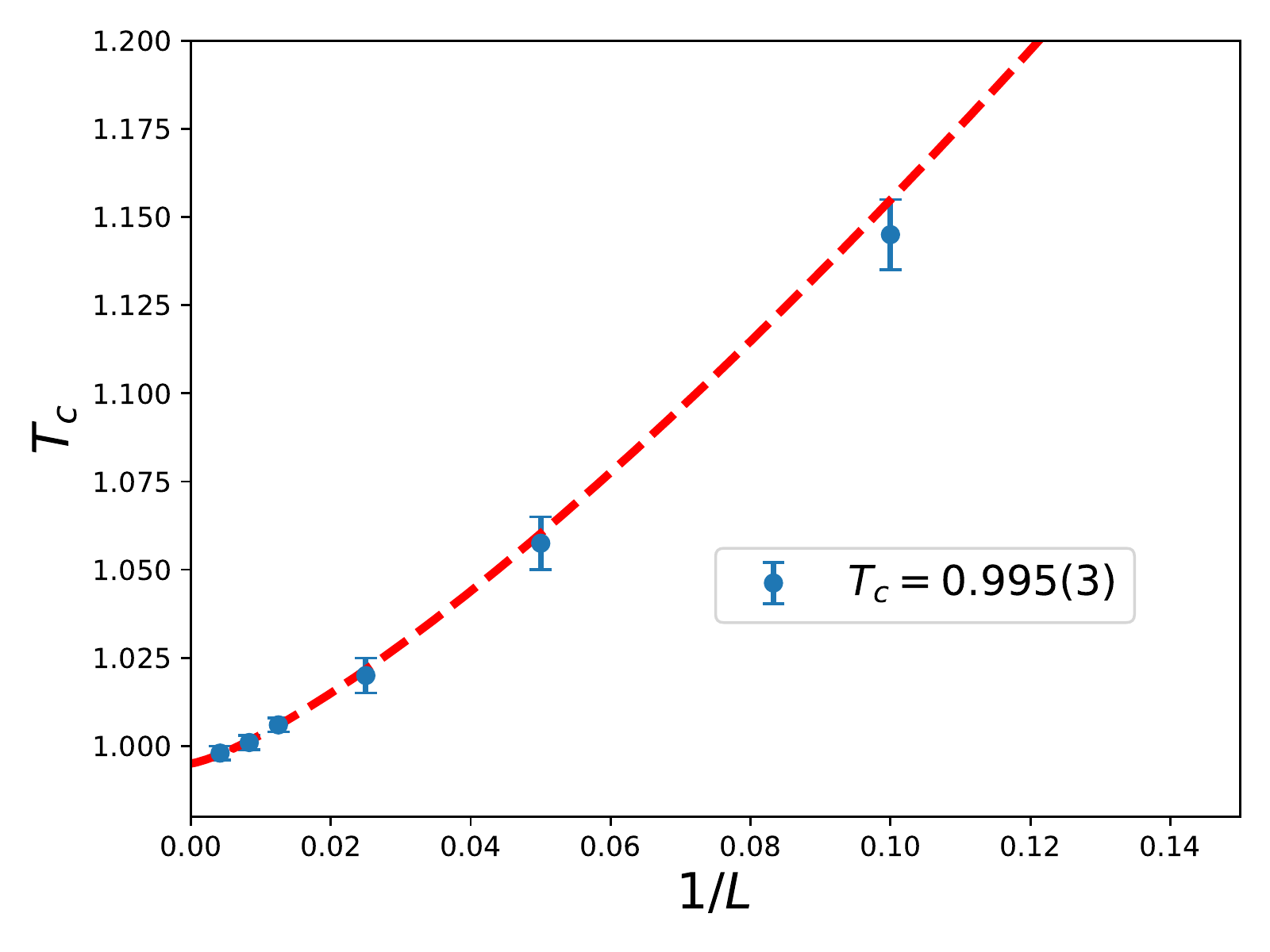}
\end{center}\vskip-0.7cm
\caption{Fit of the crossing points (of various finite $L$) to the ansatz $a+b/L^{c}$. The data
  are associated with the 3-state ferromagnetic Potts model on the square lattice \cite{Li18}
  and the dashed line in the figure is obtained by using the results of the fits.}
\label{Tc_potts}
\end{figure}

The fits lead to $a = 0.995(3)$ which agrees quantitatively with the theoretical prediction $T_c \sim 0.99497$,
see fig.~\ref{Tc_potts}.
This in turn confirms the validity of calculating the critical points using
the NN approach presented in this study. 

\begin{figure}
\begin{center}
\includegraphics[width=0.5\textwidth]{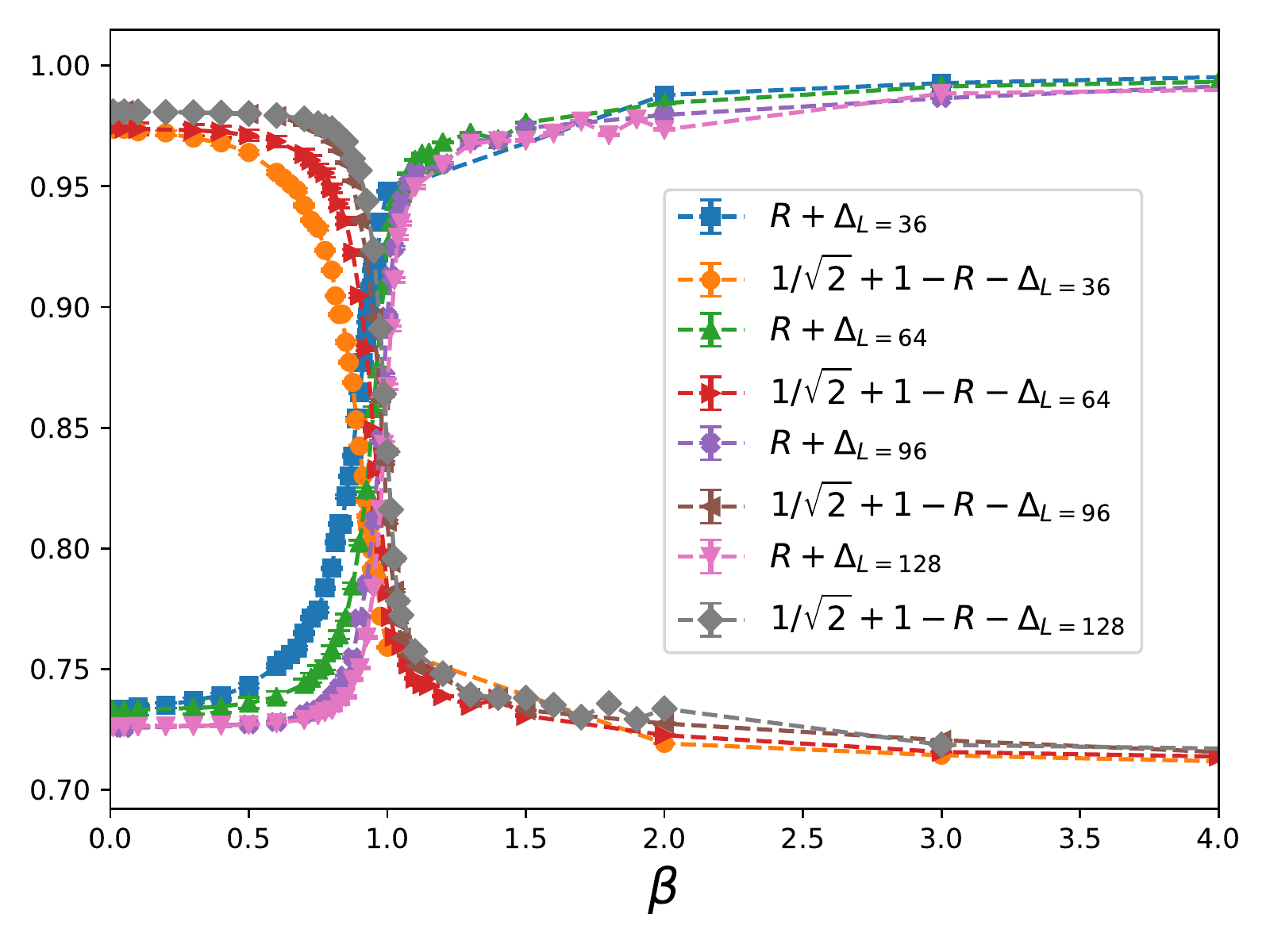}
\end{center}\vskip-0.7cm
\caption{$R+\Delta$ and $1/\sqrt{2} + 1 - R -\Delta$ as functions of $\beta$ for various $L$
  for the 2D classical XY model on the square lattice.}
\label{XYR}
\end{figure}

\begin{figure}
\begin{center}
\vbox{
~~~~~~~~~~~~~~~~~
\includegraphics[width=0.5\textwidth]{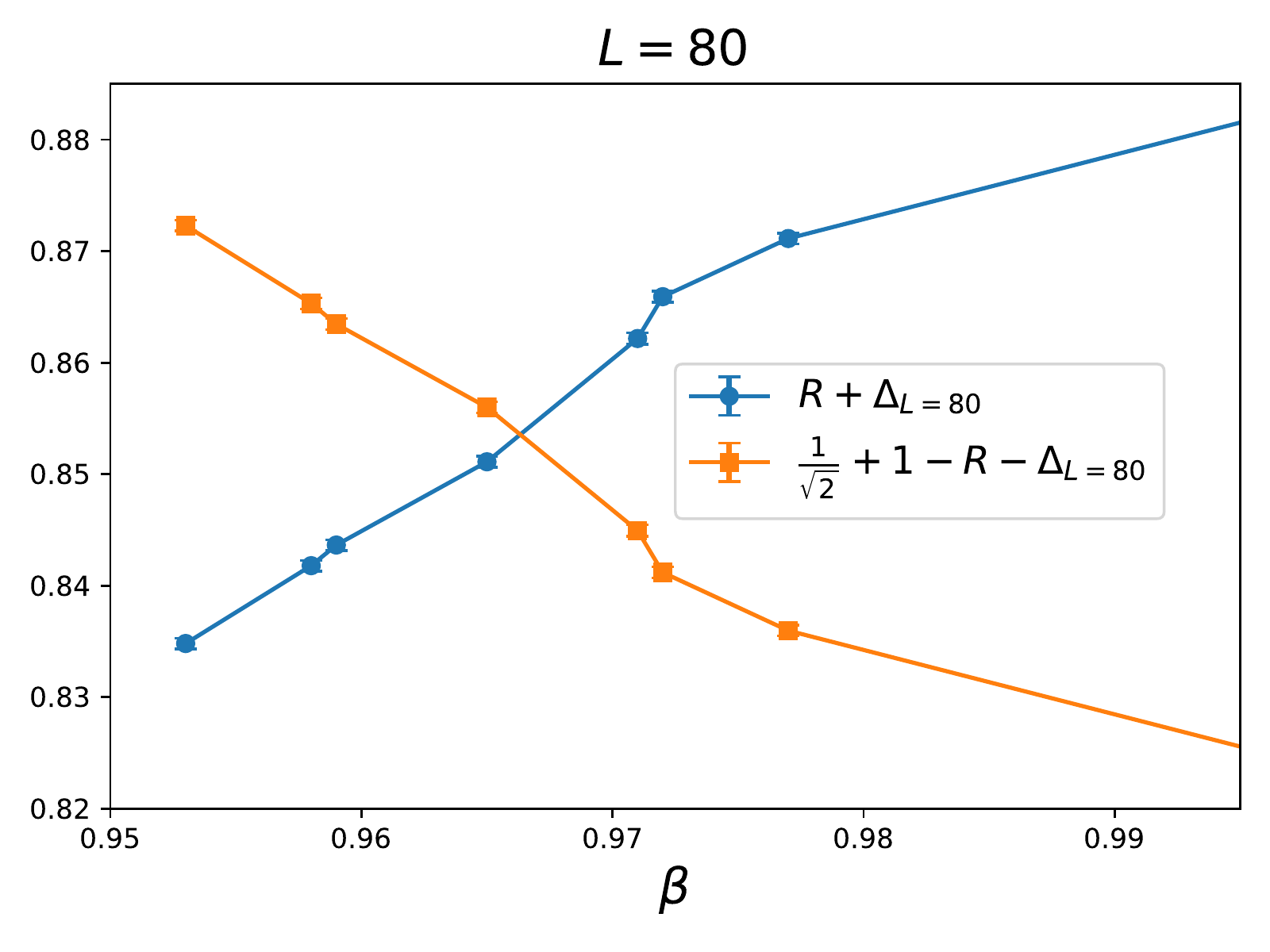}\vskip0.025cm
\includegraphics[width=0.5\textwidth]{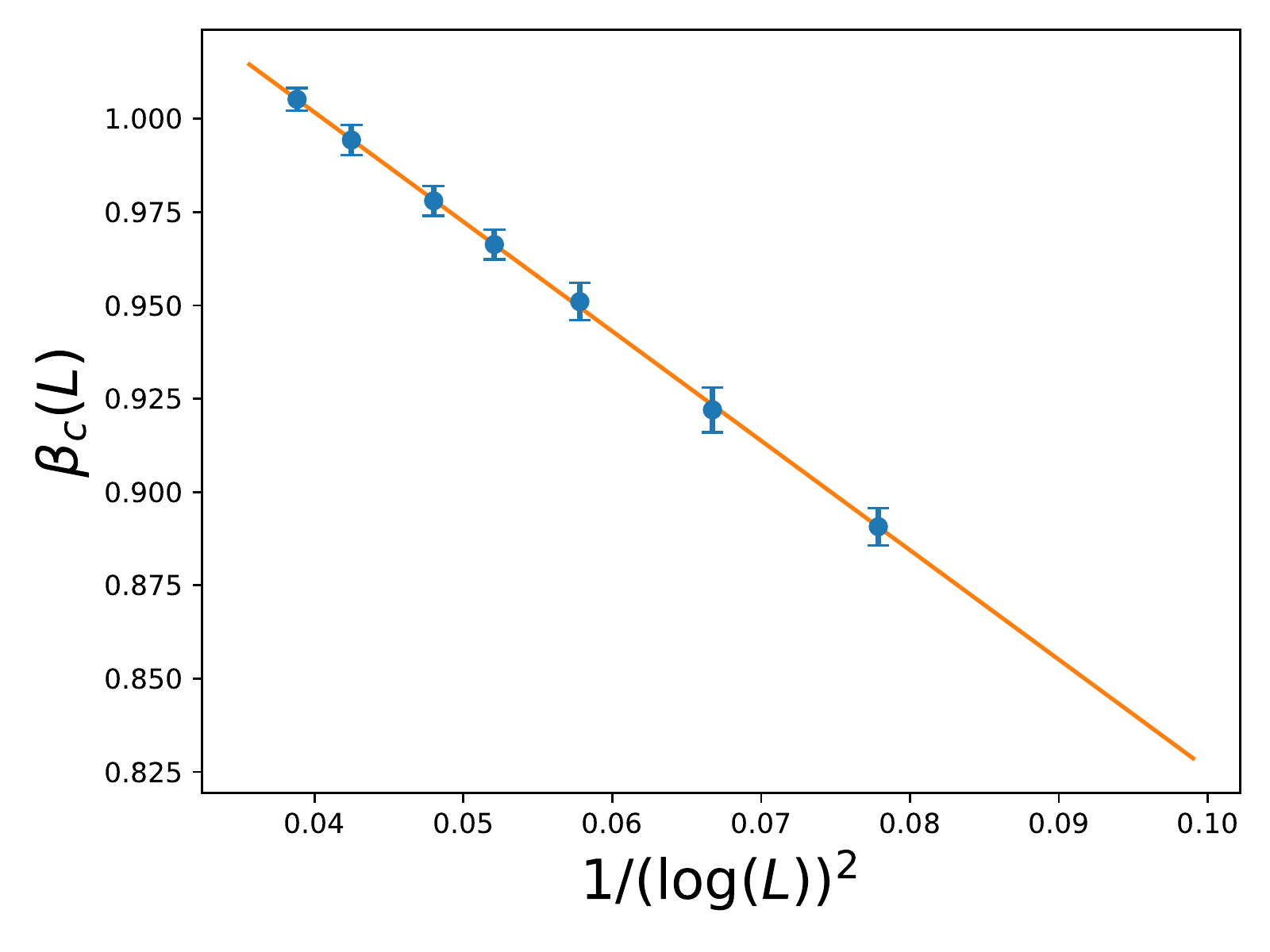}~~~~~~
}
\end{center}\vskip-0.7cm
\caption{(Top) Estimation of the crossing point for $L=80$. (Bottom) Fit of $\beta_c(L)$ to
the ansatz $a_1 + b_1/\left(\log(L)\right)^2$. The solid line is obtained using the results from the fit.}
\label{XYbetac}
\end{figure}

\subsubsection{2D classical XY model}

The $R+\Delta$ and $1/\sqrt{2} + 1 - R -\Delta$ as functions of $\beta$ for several $L$ of the 2D classical XY model are demonstrated
in fig.~\ref{XYR}. Similar to the analysis done for the 2D 3-state ferromagnetic Potts model, we would like to calculate the crossing
points for various $L$ and use some kind of finite-size scaling to fit the obtained data so that one can determine the
associated $\beta_c$ (or $T_c$). After obtaining coarse estimations of the crossing points for various $L$ from fig.~\ref{XYR},
more simulations are carried out in order to reach a better precision for these crossing points. These refined $\beta_c(L)$ are
then fitted to the same formula (i.e. $a+b/L^c$) as that used for the 3-state Potts model. We find that the obtained results are not satisfactory.
This can be expected since the topological
characteristics of the Kosterlitz-Thouless transition should reflect on $R$. 

Motivated by the finite-size scaling formulas used in Refs.~\cite{Bea18,Has05,Hsi13} for the 2D classical XY model, we use a ansatz of
the form $a_1 + b_1/(\log(L))^2$ (here $a_1$ is the $\beta_c$) to fit the newly obtained data of crossing points. The outcome is good
and we find the $\beta_c$ is given by $\beta_c = 1.119(7)$ (see both panels of
fig.~\ref{XYbetac}) which match very well with the known result
$\beta_c \sim 1.1199$ in the literature. It is intriguing that the simple NN procedure used here
works for the phase transition(s) associated with topology as well.

Before ending this subsection, we would like to point out that in principle the supervised NN method
is a optimization procedure. As a result, to obtain a more accurate estimation of the critical point in
a (supervised) NN investigation, the systematic impact associated with the tunable parameters of a built NN,
such as the number of epoch, batchsize, nodes in the hidden layers and so on, should be examined.

\section{Discussions and Conclusions}

\begin{figure}
\begin{center}
\includegraphics[width=0.5\textwidth]{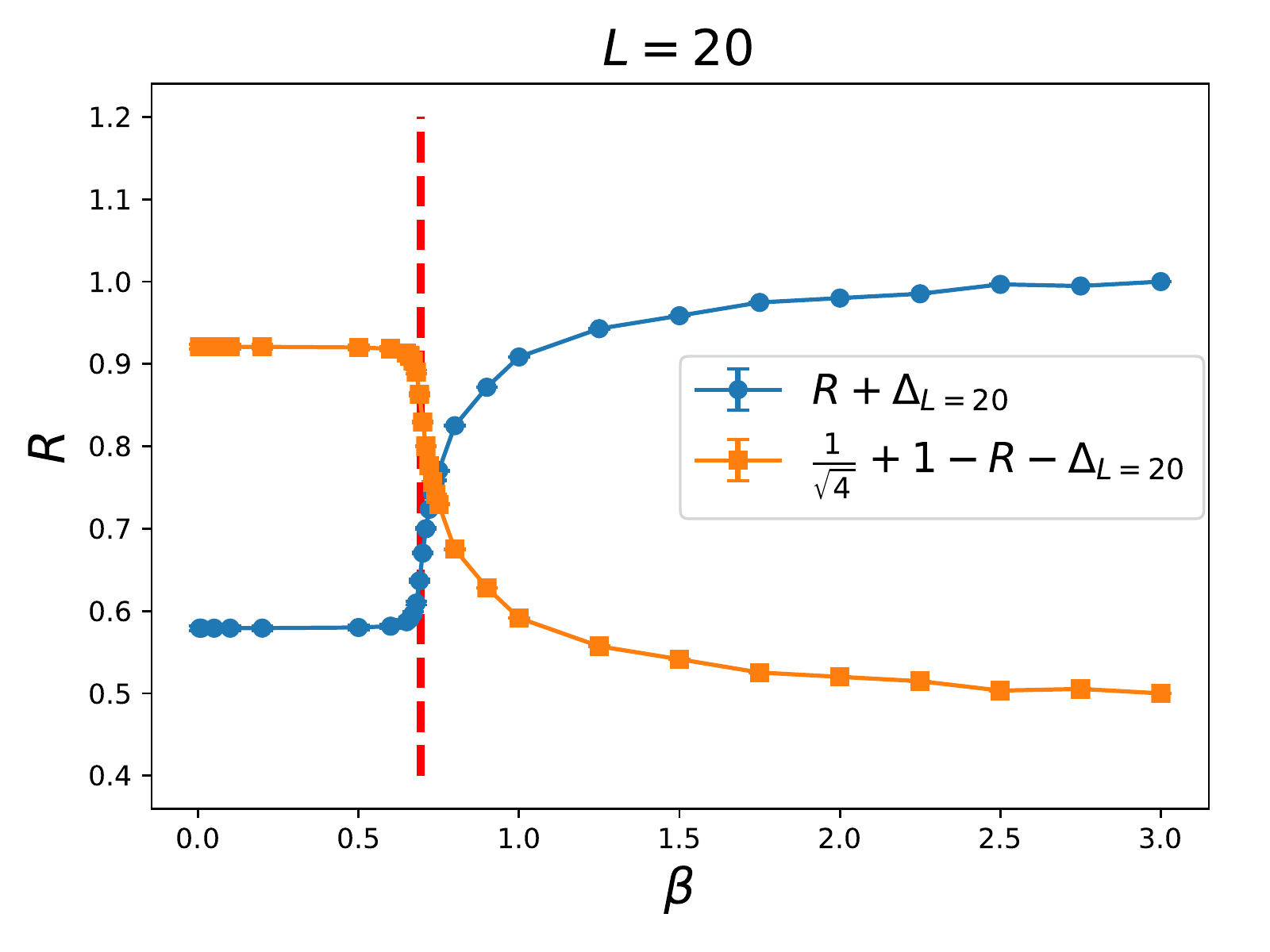}
\end{center}\vskip-0.7cm
\caption{$R+\Delta$ and $1/\sqrt{4}-\Delta + 1- R$ as functions of $\beta$ for
  the 3D classical $O(3)$ model. The results are obtained from the calculations
  which use 4 configurations as the training set.}
\label{R_4_O3_L20}
\end{figure}

In this study we investigate the phase transitions of 3D classical $O(3)$ model and
2D classical XY model, 
as well as the quantum phase transitions of both 2- and 3-D dimerized
spin-1/2 antiferromagnetic Heisenberg models using the simplest
deep learning NN, namely a MLP that is made up of only one input layer,
one hidden layer, as well as one output layer.

In our investigation, the training set for each of the studied models
consists of only two objects. In particular, none of the used
training objects belongs to the theoretical or the real configurations of
the considered physical systems.

Remarkably, with such an unconventional approach of carrying out the
training processes in conjunction with semi-experimental finite-size scaling formulas, 
the resulting outcomes from the built MLP lead
to very good estimations of the targeted critical points.
The results reached here as well as that shown in Refs.~\cite{Li18,Tan20.1}
provide convincing evidence that the performance of certain unconventional
strategies,
such as employing the theoretical ground state configurations
as the training sets, are impressive. Particularly, the simplicity of
these approaches make them cost-effective in computation.
It is amazing that the simple procedures used in Refs.~\cite{Li18,Tan20.1} and
here are not only valid for phase transitions associated with SSB, but also
work for those related to topology.

We would like to point out that for the 3D classical $O(3)$ model, the training
set used here consists of two configurations (their elements are either all 1 or all 0).
In principle, one can consider training set made up of three, four, or even five
configurations following the same idea as that of two objects training set.
To examine whether using the training sets, which constitute more than two 
objects, one can arrive at the same level of success as that shown in the
previous section, we have performed three more NN calculations using $n=3$, $n=4$,
and $n=5$ training sets. Here $n$ denotes the number of objects containing in
the training set. Interestingly, the precision of the estimated 
$T_c$ of the 3D classical $O(3)$ model obtained from these additional calculations is
becoming slightly less satisfactory with $n$, see
fig.~\ref{R_4_O3_L20} for a outcome related to $n=4$ and $L=20$. Intuitively, this can be understood as follows.
Let us assume that initially all the unit vectors 
belong to a category of the classification scheme implemented in the training stage. Then any local 
fluctuation will have greater impact on the resulting NN outputs if the training set contains more
types of objects. Despite this, it is beyond doubt that
the outcomes associated with training sets consisting of only 2 configurations, including
those from all the three studied models, strongly suggest the
effectiveness of the approach presented in this study.

\begin{figure}
\begin{center}
\vbox{
~~~~~~~~~~~~~~~~~
\includegraphics[width=0.5\textwidth]{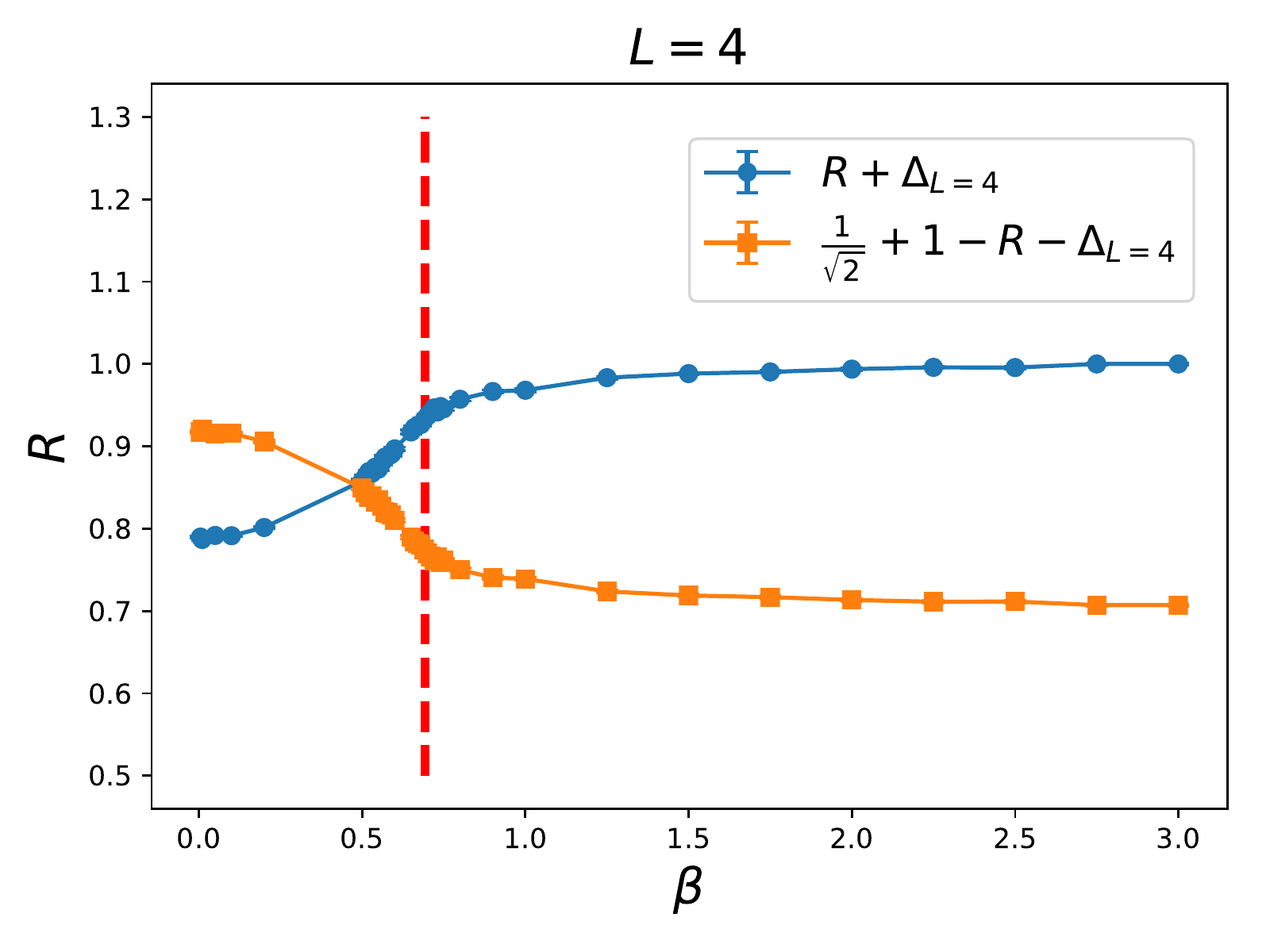}\vskip0.025cm
\includegraphics[width=0.5\textwidth]{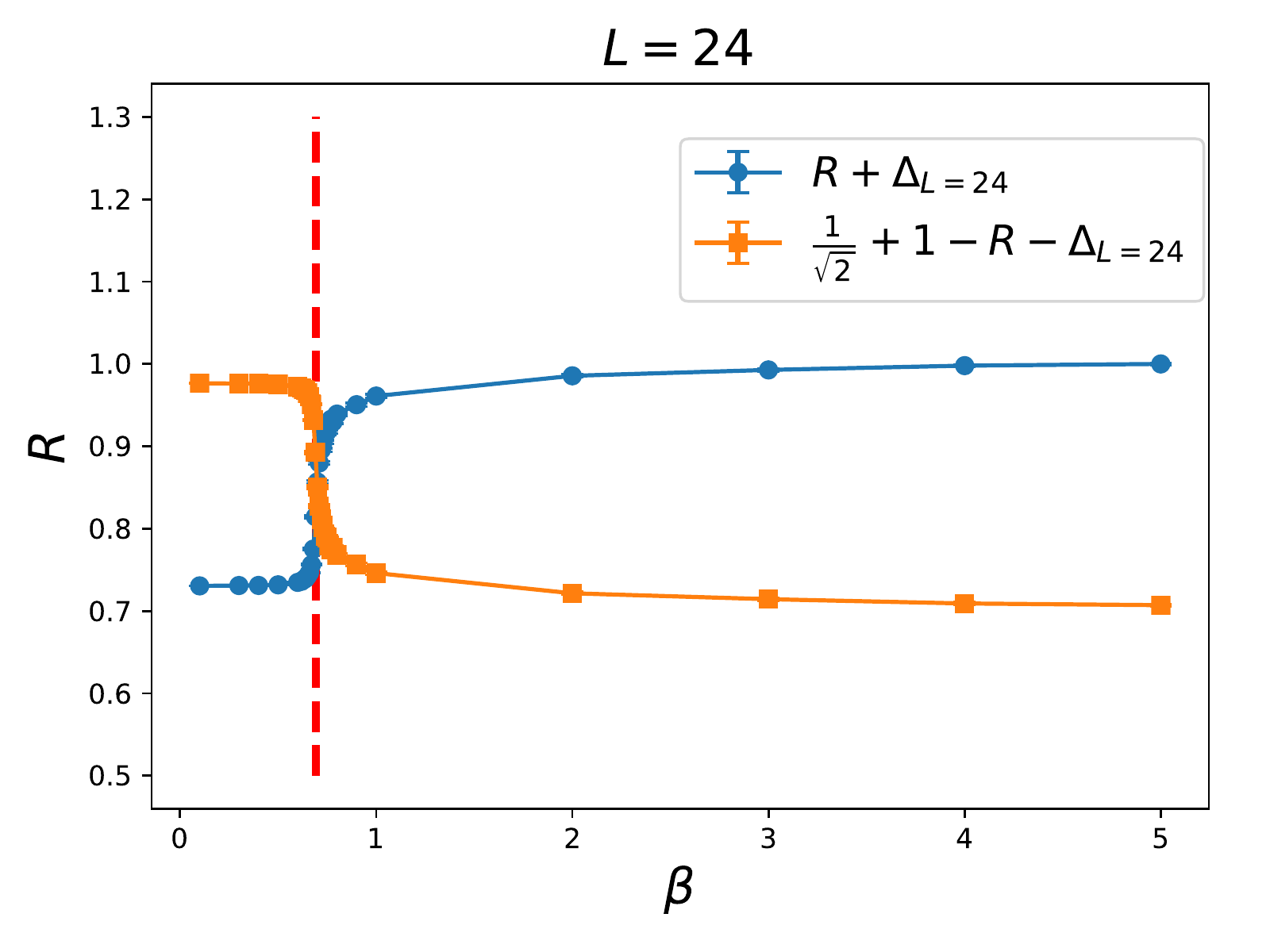}~~~~~~
}
\end{center}\vskip-0.7cm
\caption{$R+\Delta$ and $1/\sqrt{2}-\Delta + 1- R$ as functions of $\beta$ for
  the 3D classical $O(3)$ model. The data are obtained using only one set of random seeds.
  The top and bottom panels are for $L=4$ and $L=24$, respectively.}
\label{single1}
\end{figure}

\begin{figure}
\begin{center}
\includegraphics[width=0.5\textwidth]{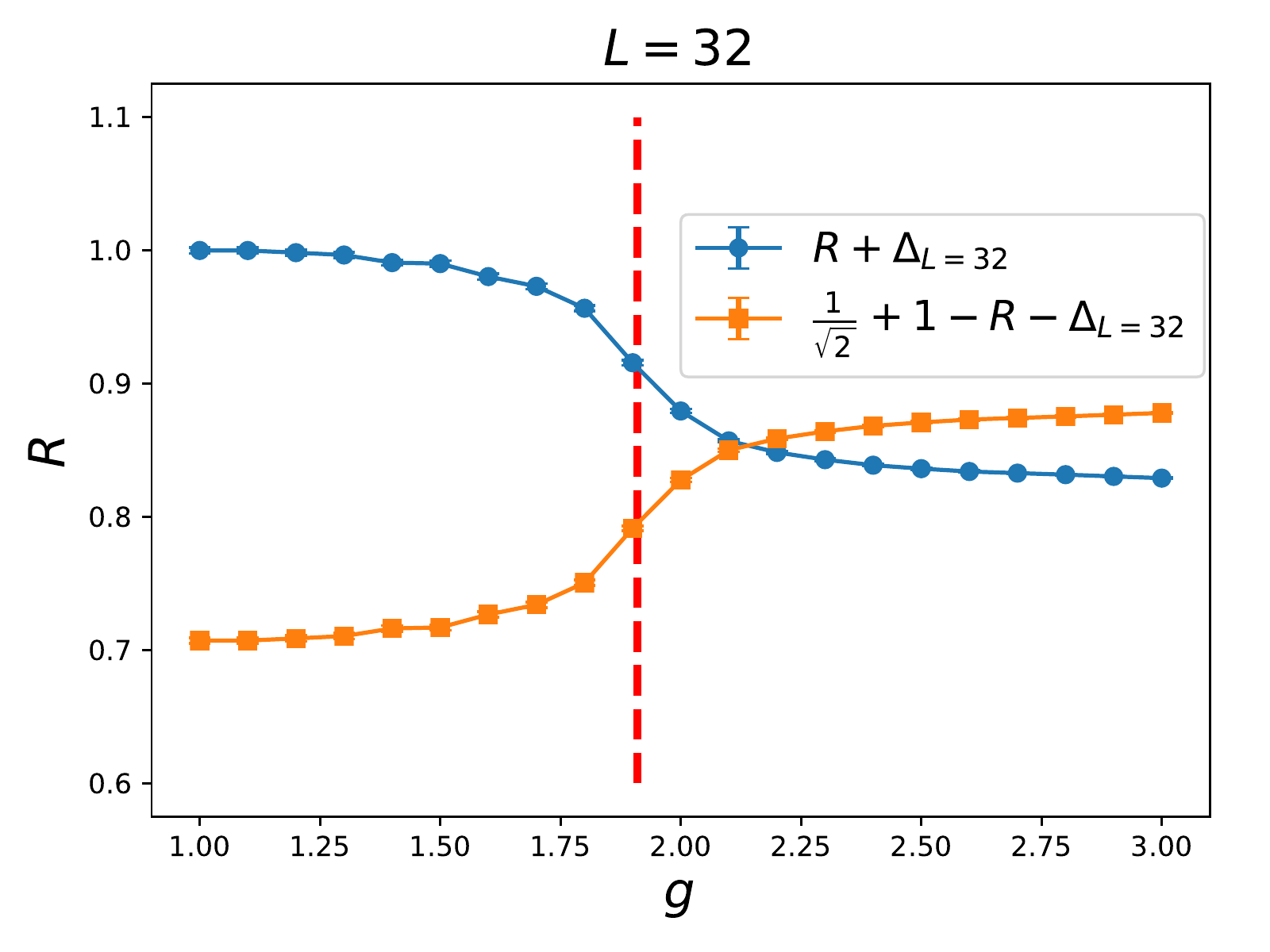}
\end{center}\vskip-0.7cm
\caption{$R+\Delta$ and $1/\sqrt{2}-\Delta + 1- R$ as a function of $g$ for
  the 2D ladder model. The data are obtained using only one set of random seeds.}
\label{single2}
\end{figure}

\begin{figure}
\begin{center}
\includegraphics[width=0.5\textwidth]{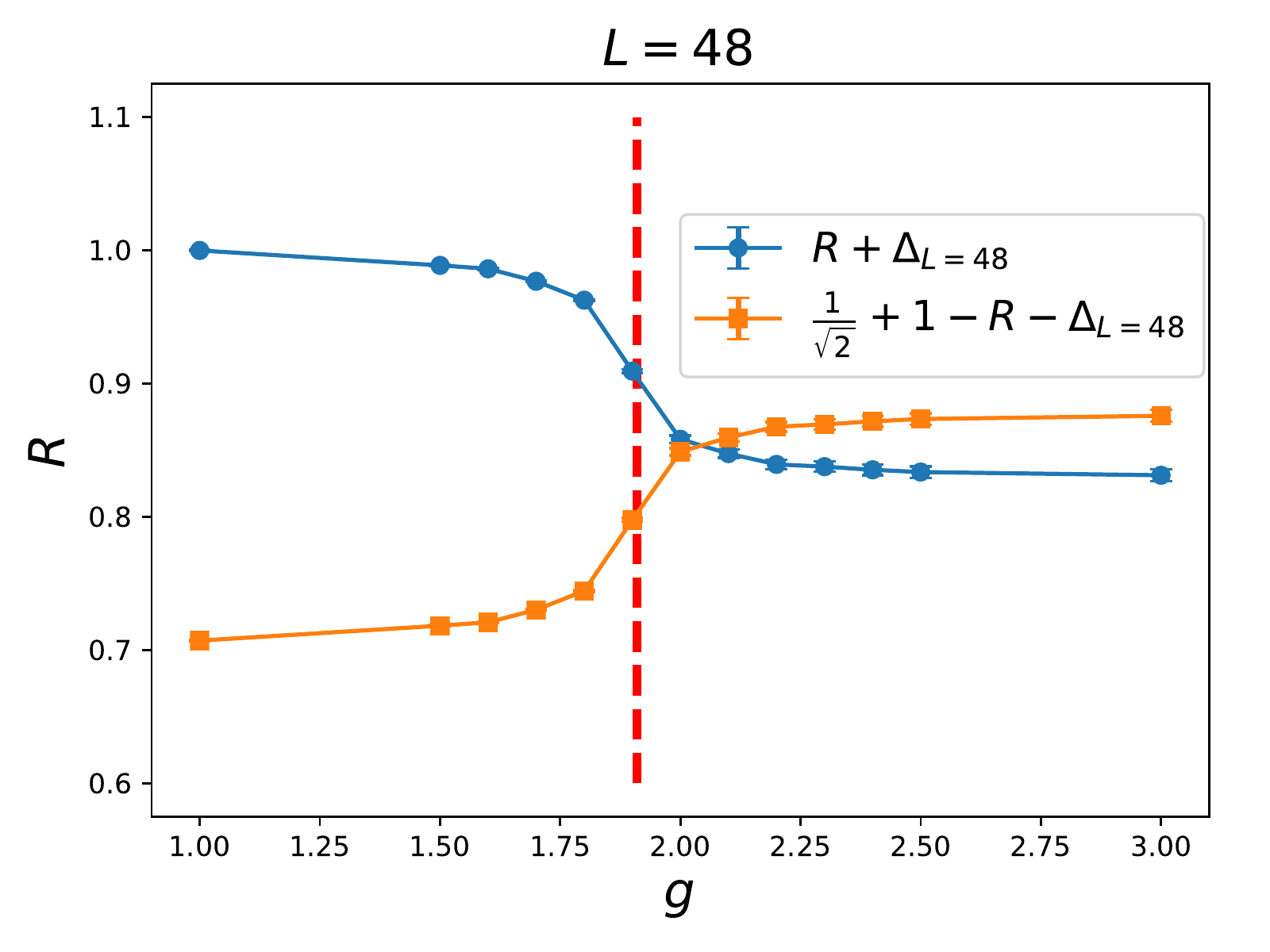}
\end{center}\vskip-0.7cm
\caption{$R+\Delta$ and $1/\sqrt{2}-\Delta + 1- R$ as a function of $g$ for
  the 2D ladder model. The data are obtained using different NN parameters from that shown
  in the previous subsection.}
\label{various}
\end{figure}

The NN results related to all the models considered here
are obtained using 10 sets of random seeds with other parameters of NN being fixed in the calculations.   
For several $L$ of the studied 3D $O(3)$ and 2D ladder models, we have performed analysis
using only one of the 10 trained NNs. Some of the resulting outcomes are shown in figs.~\ref{single1}, \ref{single2}
(The errors of the data shown in these new figures are associated with the configurations determined
from QMC simulations). These new figures match nicely with that determined with 10 sets of random seeds.
Apart from this, we have also carried out several calculations using various batchsize, epoch, and nodes in the hidden layer.
These new calculations lead to very good agreement with that shown explicitly in this study as well, see fig.~\ref{various} for
one result from these new calculations.
The additional investigations introduced in this paragraph imply that the tunable parameters of NN have very mild effects
on the resulting outcomes of $R$ for the considered
models. Hence the obtained conclusion here should be reliable. Of course, as already being pointed out before, considering other systematic
impacts are required if a highly accurate estimation of the targeted critical point is desirable.

Although in this study we have focused on studying the phase transitions of several models, it is probable
that simple NN approaches, similar to the one(s) considered here, are available for investigating
other physical properties of many-body systems.
Finally, we would like to emphasize the motivations for the series of our studies
of applying the NN techniques to investigate the phase transitions of several
physical systems, as shown in Refs.~\cite{Li18,Tan20.1} and here. Conventionally, the application of
a supervised NN to explore the critical phenomenon of a specific system has a caveat, namely
the knowledge of the critical point is required in advance before one can employ the methods of
NN for the investigation. Hence for systems with unknown critical points, it may not be easy
to apply such standard NN procedures to the studies in a straightforward manner. The approaches
considered in Refs.~\cite{Li18,Tan20.1} and here definitely can take care of this issue, hence promote
the use of NN methods in various fields of many-body systems. In particular, these unconventional methods
are adequate for carrying out any NN investigations of examining whether certain proposed theories are relevant for
a real and unexplored physical system.

\section*{Acknowledgement}
Partial support from Ministry of Science and Technology of Taiwan is 
acknowledged.


\begin{thebibliography}{1}
\bibitem{delta}
  Based on the employed training sets, the correction $\Delta$ should be calculated at the $R$ obtained with
  the largest $\beta$ for the 3D classical $O(3)$ and 2D classical XY models. Similarly, the $\Delta$ associated with the studied
  quantum spin models are determined at the $R$ corresponding to $g=1$.
\end{thebibliography}
\end{document}